\documentclass{cta-author}

\usepackage{etex}
\usepackage{array}
\usepackage{amsmath}
\usepackage{amsfonts}
\usepackage{relsize}
\usepackage{multicol}
\usepackage{multirow}
\usepackage{graphicx}
\usepackage{subfloat}
\usepackage{float}
\usepackage[margin=0.7in]{geometry}
\usepackage{setspace}
\usepackage{color,soul}
\usepackage{tikz}
\usepackage{booktabs}
\usepackage{array}
\usepackage{color}
\usepackage{nomencl}
\usepackage{amsfonts}
\usepackage{amsmath}
\usepackage[utf8]{inputenc}
\usepackage{multirow}
\usepackage{ctable}

\usepackage{tikz}
\usepackage{booktabs}
\usepackage{array}
\newcolumntype{P}{>{\fontfamily{fdm}\selectfont\small}p{3cm}}
\usepackage{color,soul}
\usepackage{subfloat}
\usepackage{subfig}
\usepackage{nomencl}
\makenomenclature
\renewcommand\nomgroup[1]{%
  \item[\bfseries
  \ifstrequal{#1}{A}{Acronyms}{%
  \ifstrequal{#1}{S}{Systems}{%
  \ifstrequal{#1}{V}{Variables}{
  \ifstrequal{#1}{C}{Constants}{}}}}%
]}
\usepackage{algorithm}
\usepackage{algpseudocode}
\usepackage{algorithmicx}
\usepackage{float}
\usepackage{changepage}
\usepackage{threeparttable}

\usepackage{subfig}
\usepackage{graphicx}
\usepackage{multirow}
\usepackage{hhline}
\usepackage{picinpar}
\usepackage{amsmath}
\usepackage{url}
\usepackage{flushend}
\usepackage{colortbl}
\usepackage{soul}
\usepackage{tikz}
\usepackage{booktabs}
\usepackage{multirow}
\usepackage{ctable}
\usepackage{relsize}
\usepackage{multicol}
\usepackage{pifont}
\usepackage{color}
\usepackage{alltt}
\usepackage{enumerate}
\usepackage{siunitx}
\usepackage{breakurl}
\usepackage{epstopdf}
\usepackage{pbox}
\usepackage{subfloat}
\usepackage{array}
\usepackage{colortbl}
\usepackage{setspace}
\usepackage{relsize}

\usepackage{environ}
\NewEnviron{myequation}{%
\begin{equation}
\scalebox{0.85}{$\BODY$}
\end{equation}
}

{}
{}
{}

\begin{document}

\title{A Case Study on the Effects of Partial Solar Eclipse on Distributed Photovoltaic Systems and Management Areas}
\author{\au{Aditya Sundararajan$^1$}, 
\au{Temitayo O. Olowu$^1$},
\au{Longfei Wei$^1$},
\au{Shahinur Rahman$^1$},
\au{Arif I. Sarwat$^{1*}$}
}

 \address{
\add{1}{Department of Electrical and Computer Engineering, Florida International University, Miami, USA}
\email{asarwat@fiu.edu}
}

\begin{abstract}
Photovoltaic (PV) systems depend on irradiance, ambient temperature and module temperature. A solar eclipse causes significant changes in these parameters, thereby impacting PV generation profile, performance, and power quality of larger grid where they connect to. This paper presents a case study to evaluate the impacts of the solar eclipse of August 21, 2017 on two real-world grid-tied PV systems ($1.4$MW and $355$kW) in Miami and Daytona, Florida, the feeders they are connected to, and the management areas they belong to. Four types of analyses are conducted to obtain a comprehensive picture of the impacts using 1-minute PV generation data, hourly weather data, real feeder parameters, and daily reliability data. These analyses include:
individual PV system performance measurement using power performance index; power quality analysis at the point of interconnection; a study on the operation of voltage regulating devices on the feeders during eclipse peak using an IEEE 8500 test case distribution feeder; and reliability study involving a multilayer perceptron framework for forecasting system reliability of the management areas. Results from this study provide a unique insight into how solar eclipses impact the behavior of PV systems and the grid, which would be of concern to electric utilities in future high penetration scenarios.
\end{abstract}

\maketitle

\renewcommand{\nompreamble}{Unless otherwise specified, the following nomenclature is applied throughout this paper.}
\nomenclature[A]{PV}{Photovoltaic}
\nomenclature[A]{RES}{Renewable energy source}
\nomenclature[A]{POI}{Point of interconnection}
\nomenclature[A]{RMS}{Root mean square}
\nomenclature[A]{PPI}{Power performance index}
\nomenclature[A]{MLP}{Multilayer perceptron}
\nomenclature[A]{VR}{Voltage regulator}
\nomenclature[A]{OLTC}{On/off load tap changer}
\nomenclature[A]{GHI}{Global horizontal irradiance}
\nomenclature[A]{DAS}{Data acquisition system}
\nomenclature[A]{SAIDI}{System average interruption duration index}
\nomenclature[A]{CAIDI}{Customer average interruption duration index}
\nomenclature[A]{SAIFI}{System average interruption frequency index}
\nomenclature[A]{MAIFI}{Momentary average interruption frequency index}
\nomenclature[A]{PR}{Performance ratio}
\nomenclature[A]{STC}{Standard test conditions}
\nomenclature[A]{THD}{Total harmonic distortion}
\nomenclature[A]{TDD}{Total demand distortion}
\nomenclature[A]{IFL}{Instantaneous flicker}
\nomenclature[A]{BP}{Backpropagation}
\nomenclature[A]{PDF}{Probability density function}
\nomenclature[V]{$P_{DC}$}{Nameplate capacity of the PV system}
\nomenclature[V]{$t$}{Time-step instance}
\nomenclature[V]{$Ir(t)$}{Irradiance measured in $W/m^2$ at time $t$}
\nomenclature[V]{$P_{estimate}(t)$}{Expected power from PV system at time $t$}
\nomenclature[V]{$E_{estimate}$}{Expected energy from PV system for a year}
\nomenclature[V]{$p_{dirt}$}{De-rate coefficient representing dirt}
\nomenclature[V]{$p_{panel}$}{Maximum wattage of the PV module ($W$)}
\nomenclature[V]{$p_{MPP}$}{Maximum power point wattage of the PV module ($W$)}
\nomenclature[V]{$p_{mismatch}$}{The PV module mismatch coefficient \\= $\frac{P_{panel}-P_{MPP}}{P_{MPP}}$}
\nomenclature[V]{$p_{cable}$}{De-rate factor capturing cable losses}
\nomenclature[V]{$p_{inverter}$}{De-rate factor capturing inverter's conversion loss, also called efficiency}
\nomenclature[V]{$D$}{Net de-rate factor of the PV system}
\nomenclature[V]{$T(t)$}{PV module's temperature at time $t$}
\nomenclature[V]{$\%_{temp\_coeff}$}{PV module's rated temperature coefficient}
\nomenclature[V]{$T_{cell\_avg}$}{Average temperature of a PV cell ($^\circ~C$)}
\nomenclature[V]{$kWhAC_{actual}$}{Energy produced by the PV system as observed ($kWh$)}
\nomenclature[V]{$kWAC_{actual}(t)$}{Power generated by the PV system at time $t$}
\nomenclature[V]{$PPI(t)$}{Power performance index of the PV system at time $t$}
\nomenclature[S]{$\mathcal{A}$}{The $1.4$MW PV system located at Miami, FL}
\nomenclature[S]{$\mathcal{B}$}{The $355$kW PV system located at Daytona, FL}
\nomenclature[V]{$Q_{cb}$}{Reactive Power Injection by Capacitor Banks}
\nomenclature[V]{$Tap_t$}{Tap Position of the LTCs at time $t$}
\nomenclature[V]{$N_{taps,t}$}{Number of tap changes made by the LTCs at time $t$}
\nomenclature[V]{$N_{max-tap,t}$}{Maximum allowable number of tap changes}
\nomenclature[V]{$N_{\text{Daily Cap Switching}}$}{Maximum allowable number of tap changes}
\nomenclature[V]{$N_{\text{Cap Switching},t}$}{Number of switching operations at time $t$}
\nomenclature[V]{$N_{\text{Cap Switching}}^{max}$}{Maximum total number of capacitor switching operations per day}
\nomenclature[V]{$V_{n,t}$}{Actual bus voltage at time t,}
\nomenclature[V]{$V_{n}^{max}$}{Maximum allowable node voltage}
\nomenclature[V]{$V_{n}^{min}$}{Minimum allowable node voltage}
\nomenclature[V]{$R$}{Feeder resistance}
\nomenclature[V]{$X$}{Feeder reactance}
\nomenclature[V]{$P_L$}{Load active power }
\nomenclature[V]{$P_{PV}$}{Injected active power}
\nomenclature[V]{$Q_L$}{Load reactive power}
\nomenclature[V]{$Q_{PV}$}{Injected reactive power}
\nomenclature[V]{$Q_{base}$}{Base voltage at POI}
\nomenclature[V]{$X_{n,RMS}$}{RMS magnitude (either voltage or current) of $n^{th}$ harmonic}
\nomenclature[V]{$X_{o,RMS}$}{RMS magnitude (either voltage or current) of fundamental frequency}
\nomenclature[V]{$I_{n,RMS}$}{RMS magnitude current of $n^{th}$ harmonic}
\nomenclature[V]{$P_{st}$}{Short-term flicker}
\nomenclature[V]{$P_{lt}$}{Long-term flicker}
\printnomenclature

\section{Introduction}  \label{sec:intro}
Power generation from RESs has become inevitable owing to the environmental impacts of generating power from conventional sources such as coal. However, the generation from RESs such as PV systems is dependent on several external parameters \textemdash solar irradiance, ambient temperature, module temperature, wind velocity (wind speed and direction), dust, cloud cover (cloud coverage area, cloud density and cloud velocity), soiling, and shading \cite{Hosenuzzaman2014,adityasysjournal,protocollevel2018,theft2017}. Some intrinsic parameters such as conversion losses and cabling losses, which can be summarized as de-rate factors, also influence the net output from PV systems. However, the most dominant of these factors are the irradiance, ambient temperature, and module temperature~\cite{iotbookchap}. One event that alters these parameters over a very short duration, thereby impacting PV generation, is the solar eclipse.

A solar eclipse occurs whenever the moon passes between the Sun and the earth, blocking the path of the solar radiation to the earth. The moon is capable of entirely blocking out the Sun, since the ratio of the diameter to the distance from the earth for both the moon and the Sun is the same. A total eclipse occurs when the moon blocks the Sun entirely, and partial otherwise. There are other types of solar eclipses too, but their discussion is beyond the scope of this paper. While a solar eclipse is accurately predictable, its impact on the smart grid has been less emphasized in the literature.

Utilities conduct pre-eclipse studies using spatial and temporal profiling to regularly monitor the generation profiles of PV and profiles of connected loads. This helps system planners better to schedule and allocate their resources to cope with the impending impact of the eclipse. Prior to the North American August 21, 2017 eclipse, a combined loss of $3.5$GW in utility scale, $1.5$GW loss in rooftop PV power generation in California, and a total loss of $5.2$GW in the whole of the United States was projected \cite{CAISO2017,wide2017}. Based on weather prediction and the projected loss of generation, many other conventional generation sources and energy storage are dispatched to compensate for the loss in generation and the change in load during the eclipse \cite{en11071782}. This in-depth pre-event analysis is, however, largely restricted to utility-scale PV systems, because there is little operational visibility on distributed PV such as rooftop solar and small commercial systems at offices, universities and buildings~\cite{nreleclipse2018}. Although not of concern now, the dynamics will change when the penetration level (the volume of PV connected relative to the amount of load) of such distributed PV systems increases drastically, wherein grid operations will be significantly impacted~\cite{nrelreportaditya,duong2018}. The steep ramp rates due to the loss of generation and the possible change in load patterns due to the eclipse could potentially impact the stability of the system. Hence, there is a need to evaluate the impacts of solar eclipse on distributed PV systems to provide a roadmap for the utilities to prepare for eclipses in the future.

To, this effect, this paper explores the impacts of the eclipse of August 21, 2017 on two distributed PV systems from an individual system-level up to the management-area-level. The presented case study considers systems located in Florida with different generation nameplate capacities: $1.4$MW (System $\mathcal{A}$ at Miami) and $355$kW (System $\mathcal{B}$ at Daytona). The study is divided into four analyses during the eclipse peak: performance measurement and relationship analysis for the systems, power quality analysis at the POI of System $\mathcal{A}$, simulation of voltage device operations on an IEEE $8500$ test case distribution feeder network remodeled to include real system parameters of the feeders that the two PV systems connect to, and finally, system reliability evaluation and forecasting for the utility management areas (Miami and Daytona). Different datasets are used for conducting these analyses after being subject to quality checks~\cite{adityampce,adityaacm}: real time-series PV generation data of 1-minute resolution collected from cloud-based on-site data acquisition systems for system performance; real time-series average RMS voltage, harmonics (voltage, current) and IFL, short-term flicker ($P_{st}$), and long-term flicker ($P_{lt}$) data of 1-minute resolution collected from the POI of System $\mathcal{A}$ using a meter for power quality analysis; smart inverter data sheets (for power-efficiency curves), PV data sheets (for temperature-efficiency curve) and load profile data for the voltage profile analysis; and, hourly weather data from the National Climatic Data Center and daily reliability data for the two management areas for the reliability analysis.

The key contributions of this paper are that it: \textbf{(1)} investigates events such as solar eclipse that have short-term but high-magnitude impacts on PV generation unlike a majority of the literature that deals with fluctuations over longer periods of time (Sections \ref{sec:related} and \ref{sec:background}); \textbf{(2)} analyzes the impacts of solar eclipse on PV systems: performance (Section \ref{subsec:objective1}), power quality at POI (Section \ref{subsec:objective3}), voltage profiles of feeders (Section \ref{subsec:objective4,5}), and reliability of management areas (Section \ref{subsec:objective2}); \textbf{(3)} provides a roadmap for utility distribution planners to better handle the impacts of solar eclipses under future high PV penetration scenarios (Section \ref{sec:model}); \textbf{(4)} proposes PPI, an effective metric, to quantify instantaneous PV performance during the eclipse peak that can be used by utilities in planning studies (Section \ref{subsec:objective1r}); \textbf{(5)} quantifies power quality parameters and measures deviations from the standard values during the eclipse at existing and future high penetration scenarios to help utilities take some proactive steps to mitigate the possible power quality violations that could arise as a consequence of the eclipse event. (Section \ref{subsec:objective3r}); \textbf{(6)} quantifies the eclipse's impact on voltage regulating devices using real system parameters to enable utility companies take proactive voltage control steps through the use of smart inverter settings and optimal coordination of other voltage regulating devices in the network (Section \ref{subsec:objective4,5r}); and \textbf{(7)} develops regression models to analyze the relationship between weather parameters and reliability indices of management areas and forecast the indices, thereby helping utilities evaluate how solar eclipses impact the stability of the grid at a larger scale (Section \ref{subsec:objective2r}). Finally, Section \ref{sec:conclusion} concludes the study and provides future directions for research in the area.

\vspace{-0.5cm}
\section{Related Work} \label{sec:related}

Studies have been conducted to determine the overall behavior of PV systems during an eclipse. A study monitored the performance of a $4.85 kW$ PV system during the August 21, 2017 eclipse and estimated the performance measurements using irradiance calculation approaches~\cite{pvperf01}. However, it does not quantify the performance of the system using one of the standard accepted metrics such as performance ratio, energy performance index or PPI as recognized by the industry~\cite{pvperfnrel1,pvperfsunspec,pvperfnrel2,pvperfnrel3}. A similar study of PV performance was conducted by other authors too, but they primarily relied on comparing the net PV generation on the day of the eclipse versus the following: generation of the same system on the same date of the previous year or generation of the same system on the date prior to or next to the date of the eclipse~\cite{pvperf4,pvperf5}. While these methods provide a visual idea about the impact of the eclipse, they do not quantify the impacts as a measurable metric. Further, these studies limit their scope to a single system of concern, thus not considering potential understandings of how an eclipse could impact PV over a larger area and what that might mean for aggregation-related studies in the future. The performance metrics defined in Section \ref{subsec:objective1} are derived from industry-accepted metrics that have recently gone beyond the traditionally used performance ratio~\cite{pvperfsunspec,pvperf6}. A paper in~\cite{pvperf7} evaluated different metrics for PV performance, but it considers metrics which have certain limitations. For example, it looks at yield and capacity factor which depend on the PV system nameplate capacity, performance ratio which depends on the PV system model and the local weather parameters,  not consider PPI which is a more effective metric to compare performances of systems of different sizes, and system generation which is not a direct measure of performance.

Many research papers have studied the technical limitations and unfavorable effects on power quality such as feeder voltage variations, small range voltage fluctuations, and harmonic injections on power system parameters resulting from intermittent PV generation~\cite{HAQUE20161195, 7856217}. None of them, however, focus on special events such as solar eclipse. Voltage sag, swell, and small range voltage fluctuations are very common and could have been observed at the POI during~\cite{VoltageIssue}. The semiconductor devices used in inverter based PV systems inject significant harmonics and could increase the power losses on the grid ~\cite{6939147}. It is necessary to investigate the effects of grid-tied PV system due to different weather related events such as solar eclipse for reliable and continuous power supply. The extent of power quality impact depends on network configuration, weather variability, and the location of PV plant. The power quality analyses was carried out on System $\mathcal{A}$ at Miami.

The integration of PV systems on distribution feeders (depending on their locations) usually have some impacts on the operations on the voltage regulating devices such as VRs, OLTCs, capacitor banks, and voltage regulators \cite{Thesis2013}. These impacts are usually observed in terms of the number of switching operations of these devices. Due to PV integration, there could be an increase or decrease in the number of switching of these devices depending on the feeder profile, the location of the PV, and the operation of the smart inverters. The life span of these devices usually impacted by the number of their switching operations. This could attract some huge financial cost for utility companies due to the possible need to maintain of totally replace these devices. Most studies reported on the impact of the eclipse events did now show their impacts on these voltage regulating devices \cite{Veda2018,WideArea}. Usually, these switchings are difficult to capture on a live distribution system during their operations. The planning departments of many utility companies, therefore, simulate these events for a study-impact analysis which is usually not made public.

Severe weather conditions, such as typhoons, ice storms, and earthquakes have recently been considered to study the power system reliability performance~\cite{J1,J2,J3}. In \cite{J1}, the effects of different types of severe weather conditions on the reliability performance of power system components were evaluated, and the existing methodologies for modeling these effects were listed and compared. A reliability assessment framework was proposed in \cite{J2} for quantifying the power transmission system performance under the typhoon weather. In \cite{J3}, a defensive islanding-based operational enhancement approach was developed to improve the power system resilience to extreme weather events. This is in contrast with solar eclipse that only involves predictable changes in common weather parameters such as temperature and solar radiation. Hence, the required methods of analyses are applicable more for analyzing system reliability under common weather conditions, not extreme. In \cite{Arif1}, a power distribution system reliability assessment framework was presented using time-series common weather data. A statistical model was introduced in \cite{Arif2} to predict the daily number of common weather-based power interruptions in power distribution systems, but there is a lack of literature which explores system reliability during eclipses.

\section{Background for the Case Study} \label{sec:background}
This section provides a brief background for the case study, including some information on the August 21, 2017 eclipse.

\subsection{The Eclipse of August 21, 2017} \label{subsec:event}
The total solar eclipse of August 21, 2017 was the first to be observed in twenty-six years from the USA. It was first observed in Oregon at 10:15 AM (Pacific Time) and last observed in North Carolina at 2:49 PM (Eastern Time). During the short period of the eclipse at each location, the utilities were reported to have taken their PV systems offline, wherein a surge in load was also expected. In the State of Florida, the eclipse was only partial. The two utility management areas considered in this paper, Miami and Daytona, experienced an average coverage of about $80\%$ and $89\%$ respectively, as illustrated in Fig. \ref{fig:eclipse}. As noted in Section \ref{sec:intro}, the two systems, the feeder model and the management areas considered for the case study are described in the following subsections:

\begin{figure}[t!]
    \centering
\includegraphics[width=2in, height=1.75in]{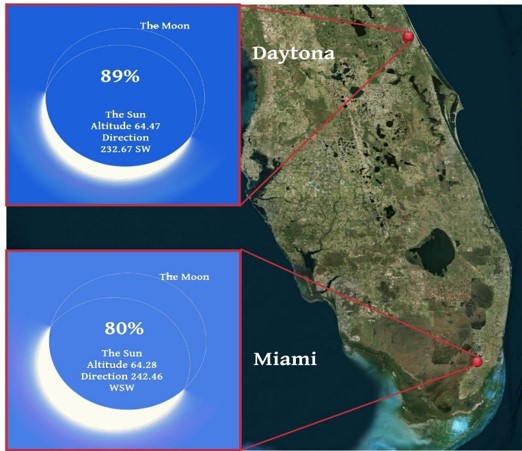}
    \caption{The statistics of the partial solar eclipse at the 2 locations.}
    \label{fig:eclipse}
\end{figure}

\begin{figure}[t!]
    \centering
\includegraphics[width=8cm,height =6cm ]{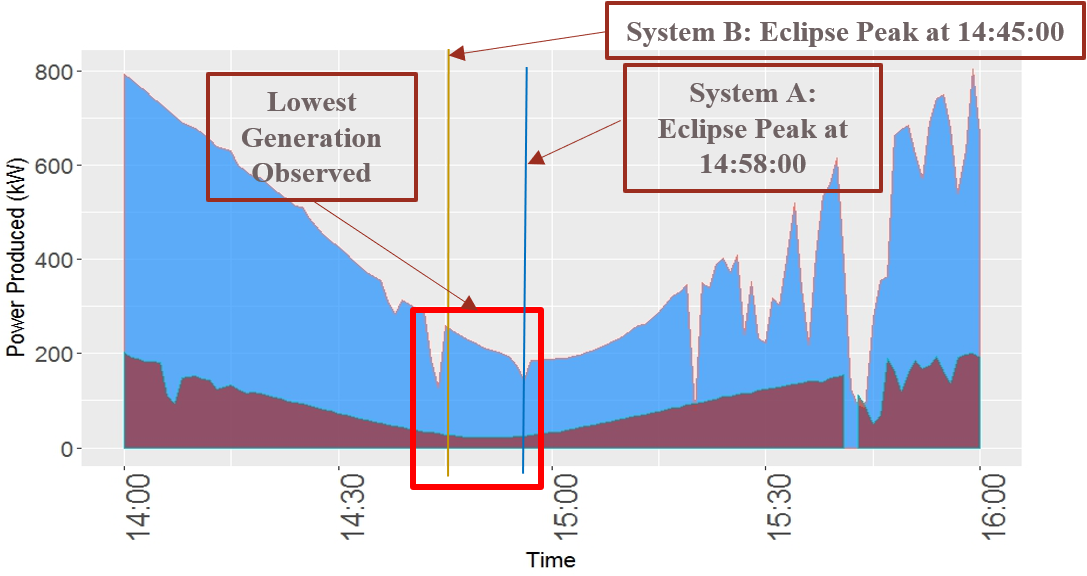}
    \caption{Total Demand Distortion during Solar Eclipse Period.}
    \label{TDD}
\end{figure}


\begin{table}[t!]
\renewcommand{\arraystretch}{1.3}
\caption{Statistics about the Partial Solar Eclipse}
\label{tab:eclipsestats}
\centering
\begin{tabular}{| >{\centering\arraybackslash}p{3cm}| >{\centering\arraybackslash}p{2cm}| >{\centering\arraybackslash}p{2cm}|}
\hline
\textbf{Value (from 2 PM to 3 PM)} & \textbf{System $\mathcal{A}$} & \textbf{System $\mathcal{B}$}\\\hline
Location & Miami & Daytona\\\hline
Time at Lowest Reading & 3:00 PM & 2:45 PM\\\hline
Drop in Power ($kW$) & 503.4 (70.8\%) & 140.64 (84\%) \\\hline
Drop in Irradiance ($W/m^2$) & 469.99 (70.8\%) & 56.3 (87.4\%)\\\hline
Drop in Ambient Temperature ($F$) & 3.48 (3.8\%) & 6.92 (7.4\%)\\\hline
Drop in Module Temperature ($F$) & 13.52 (13.1\%) & 25.4 (21.8\%)\\\hline
Change in Power Performance Index (PPI) & 1.5\% & 0.3\%\\\hline
\end{tabular}
\end{table}

\subsection{The Two Grid-Tied PV Systems $\mathcal{A}$ and $\mathcal{B}$} \label{subsec:pvsystems}
Table \ref{tab:eclipsestats} summarizes the key changes observed in net generation, average irradiance, average temperature and average PPI of the two PV systems under consideration. It can be observed that System $\mathcal{B}$, for reasons described in Section \ref{subsec:pvsystems}, experienced a greater fluctuation in module temperature, ambient temperature, and irradiance, but recorded a lower fluctuation in its instantaneous performance which is measured using the PPI. This metric is discussed in detail in Section \ref{subsec:objective1}.

The two PV systems deploy different smart inverter topologies. While System $\mathcal{A}$ uses string inverters and a cluster controller to aggregate individual inverter productions, System $\mathcal{B}$ uses a combination of micro-inverters and string inverters~\cite{pvsys1}. Locally installed weather stations measure GHI, module temperature and ambient temperature with up to a resolution of 1 minute. A cloud-based DAS is used to access the raw data for further processing and analysis, as described in the authors' previous works~\cite{missingdata,pvsys2,fogbookchap}. Systems $\mathcal{A}$ and $\mathcal{B}$ are connected to feeders $\mathcal{A}$ and $\mathcal{B}$, respectively, which are radially distributed from their substations located in Miami and Daytona, respectively.

\subsection{The IEEE 8500 Test Feeder for Systems $\mathcal{A}$ and $\mathcal{B}$} \label{subsec:testfeeder}
The standard IEEE $8500$ test network was developed from a real distribution feeder in the US. The feeder has both the medium and low voltage levels with the longest node being approximately $17 km$ from the substation. It has four capacitor banks (three controlled and one fixed), three voltage regulators with tap-changeable substation transformer. The feeder model also contains both balanced and unbalanced loads~\cite{Arritt2010}. For this simulation, the line characteristics and load profiles of the feeders $\mathcal{A}$ and $\mathcal{B}$ were used to modify the standard IEEE test feeder. The two PV systems ($\mathcal{A}$ and $\mathcal{B}$) were modeled using OpenDSS and subsequently integrated into the test feeder. The irradiance and temperature profiles during the eclipse were also used for the PV systems modeling. The penetration level of the PV on this feeder is approximately 37\%, which is defined as the ratio of the PV installed name plate capacity to the total size of the loads on the feeder.

\subsection{Management Areas Comprising Systems $\mathcal{A}$ and $\mathcal{B}$}
The Systems $\mathcal{A}$ and $\mathcal{B}$ are respectively located at the Miami and Daytona management areas. For these two areas, the reliability indices collected contain the sustained and momentary interruption events: SAIDI, CAIDI, SAIFI, CAIFI, and MAIFI. The numbers of sustained and momentary interruption events play a key point in reliability analysis, and other reliability metrics can be calculated based on these values. Therefore, the sustained and momentary interruption events happenning during the solar eclipse are selected for reliability analysis. The common weather parameters such as {temperature}, {precipitation}, {air pressure}, and {wind speed} are mainly collected from the Miami and Daytona International Airports. Additionally, the lightning data are provided by the control center of the local electric utility, which installs its own weather stations in its different management areas.

\begin{figure*}[t!]
    \centering
\includegraphics[width=15cm,height = 7cm ]{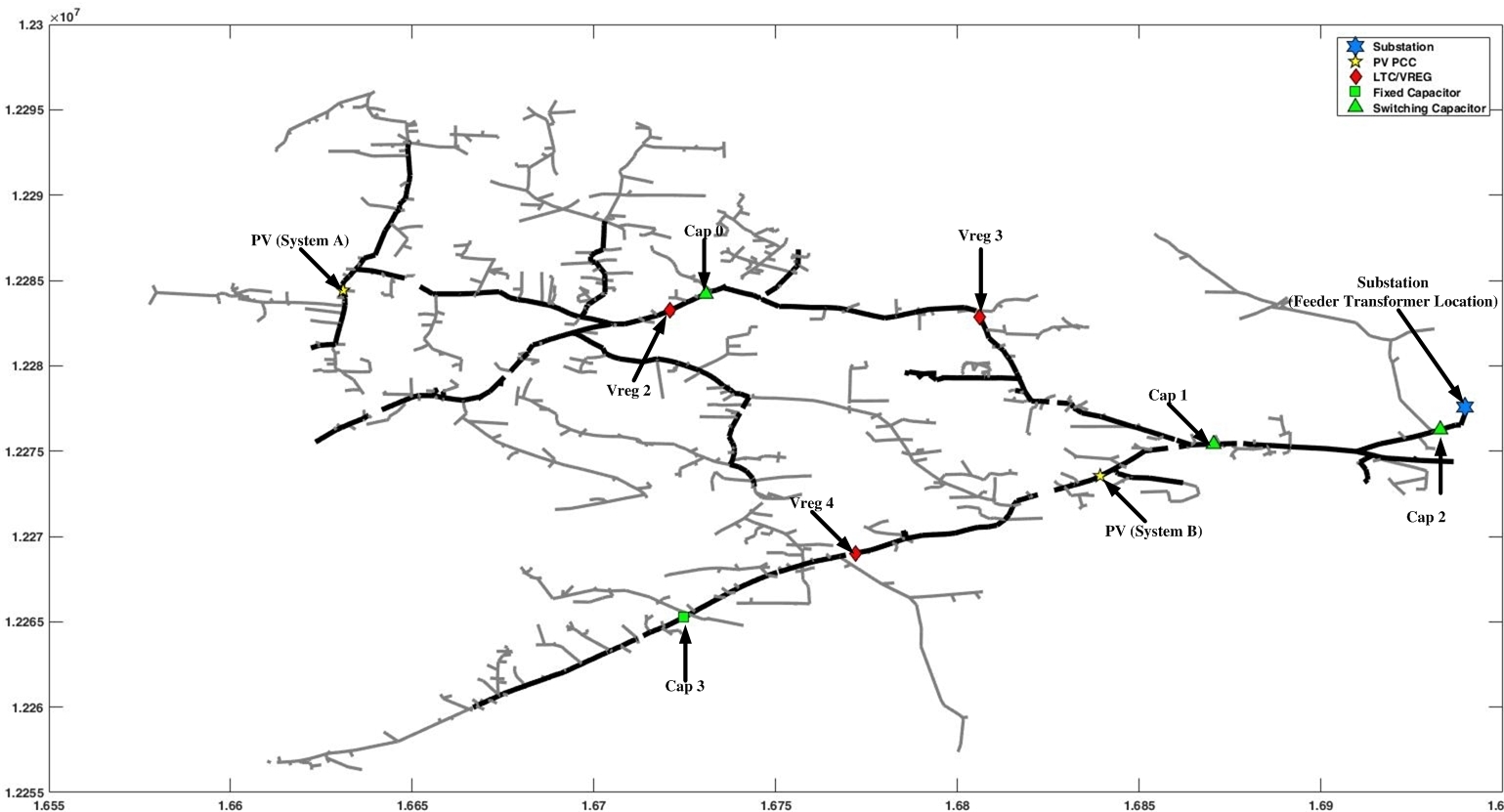}
    \caption{IEEE 8500 Test Feeder with the integration of PVs (Systems $\mathcal{A}$ and $\mathcal{B}$)}.
    \label{fig:feeder}
\end{figure*}

\section{Model Formulation} \label{sec:model}
This section describes in detail the different methods used to evaluate system performance, power quality at the POI, voltage device operations at the feeders, and management area reliability.

\subsection{PV System Performance} \label{subsec:objective1}
Many metrics are currently used to evaluate PV system performance~\cite{performancemetrics,performancemetrics1,performancemetrics2,performancemetrics3}. PR is a widely used metric, defined as the ratio of observed power (over short durations from minutes to a day) or energy (over long duration from months to a year) to the expected power or energy, respectively. Given $P_{DC}$, $Ir(t)$, and the irradiance measured at STC ($1000 W/m^2$), the values of $P_{estimate}(t)$ and $E_{estimate}$ are calculated using (\ref{eq:exppowen}), considering $1$-minute resolution data~\cite{performancemetrics}:

\vspace{-0.5cm}
\begin{center}
\begin{equation}\label{eq:exppowen}
\begin{split}
P_{estimate}(t)=P_{DC}\times\frac{Ir(t)}{1000}\times~X\times~D\\
E_{estimate}={P_{DC}\times~X\times~D\times\sum_{i=1}^{525,600} \frac{Ir(t)}{1000}}
\end{split}
\end{equation}
\end{center}

where, $D$ is a function of $p_{dirt}$, $p_{mismatch}$, $p_{cable}$, and $p_{inverter}$ such that~\cite{derate,derate1}:

\vspace{-0.3cm}
\begin{equation}\label{eq:derate}
D=p_{dirt}\times~p_{mismatch}\times~p_{cable}\times~p_{inverter}
\end{equation}

The variable $X$ in (\ref{eq:exppowen}) can be modified to improve the accuracy of the estimation and takes two parameters into account: $T(t)$ and $\%_{temp\_coeff}$. When $X=1$, it is called \textit{Uncorrected estimation}, and when $X=1+\frac{\%_{temp\_coeff}}{100}[T(t)-25]$, the estimation considers the effect of $T(t)$ by correcting it to the STC ($25^\circ~C$). It has been shown that the accuracy of estimation is maximized when $T(t)$ is corrected instead to $T_{cell\_avg}$ and wind speed~\cite{correctedpr}. In this paper, $T(t)$ is corrected just to $T_{cell\_avg}$, with all 1-minute data points averaged over a period from January $01$ through December $31$, $2017$ for the study.

\begin{table}[t!]
\renewcommand{\arraystretch}{1.3}
\caption{Power and Energy Estimation Parameters for $\mathcal{A}$ and $\mathcal{B}$}
\label{tab:ppiprofiles}
\centering
\begin{tabular}{| >{\centering\arraybackslash}p{3cm}| >{\centering\arraybackslash}p{2cm}| >{\centering\arraybackslash}p{2cm}|}
\hline
\textbf{Parameter} & \textbf{System $\mathcal{A}$} & \textbf{System $\mathcal{B}$}\\\hline
$p_{dirt}$ & 0.9 & 0.9\\\hline
$p_{mismatch}$ & 0.97 & 0.97\\\hline
$p_{cable}$ & 0.99 & 0.99\\\hline
$p_{inverter}$ & 0.98 & 0.9725\\\hline
$\%_{temp\_coeff}$ & -0.5 & -0.5\\\hline
\end{tabular}
\end{table}

PR, measured using (\ref{eq:pr})~\cite{performancemetrics2,performancemetrics3}, is widely used by the utilities to measure the performance of a particular PV system, it has some key demerits~\cite{correctedpr1}: (1) highly dependent on local weather (especially module temperature) and hence varies significantly over the course of a year; and (2) varies depending on the nameplate capacity of the system. Due to these reasons, PR is not an effective metric to compare performance of any two given PV systems.

\begin{equation}\label{eq:pr}
PR= \frac{kWhAC_{actual}}{P_{DC}}\times \frac{1000}{\sum_t Ir(t)}
\end{equation}

There exist other metrics like yield (PV systems of different sizes are not directly comparable), and PR corrected to $T(t)$ and wind speed (PV systems employing different PV models are not directly comparable). A new metric called the PPI is used in this paper, which is calculated as~\cite{correctedpr2}:

\begin{equation}\label{eq:ppi}
PPI(t) = \frac{kWAC_{actual}(t)}{P_{estimate}(t)}
\end{equation}

PPI is better to compare the performance of different PV systems because: (1) it corrects the estimation to average module temperature to account for local variations; and (2) it is independent of the inherent PV model.
There exists the energy performance index too, but it is used for comparing the performance of PV systems over an aggregated period of time. Since the eclipse lasts for a shorter period and its impacts should be evaluated over a short duration, PPI emerges as a better metric overall.

\subsection{Power Quality Analysis at the POI} \label{subsec:objective3}
Integration of inverter-based PV system into the grid introduces non-sinusoidal waveforms that degrade power quality at the POI. Indices such as average RMS voltage, THD of voltage and current, TDD, and flicker- IFL, $P_{st}$, and $P_{lt}$- are used to quantify the quality of power. Several standards, such as, IEEE Standard 1547-2018 (voltage regulation), IEEE Standard 519-2014 (harmonics regulation), and IEEE Standard 1453-2015 (flicker regulation) provide a set of criteria and requirements for the interconnection of RESs into the power grid and help in quantifying deviations from the ideal values.

PV integration could lead to changes in feeder voltage as well as the line loading . The voltage change at the POI can be expressed as follows~\cite{CHAUDHARY20183279}:

\begin{equation}
\Delta V = \frac{R (P_L-P_{PV})+ X (Q_L-Q_{PV})}{V_{base}}
\label{Voltage Change}
\end{equation}


THD is a comparison of total harmonic content in a voltage or current waveform to the $60Hz$ fundamental magnitude that is expressed in (\ref{THD})~\cite{1242609}. Typical voltage THD level varies between $0.5\%$ and $5\%$, and current THD ranges from $0.5\%$ to $80\%$ or more, depending on the type of load, amount of PV generation, PV penetration level, etc.

\begin{equation}
THD = \frac{\sqrt[]{\sum_{n=2}^\infty X_{n,RMS}^2}}{X_{o,RMS}} \times 100\%
\label{THD}
\end{equation}

where $X_{fundRMS}$ and $X_{nRMS}$ are the RMS magnitude (either voltage or current) of fundamental frequency and $n^{th}$ harmonic, respectively.

Current THD depends on the change of the RMS current magnitude without considering load types (harmonic or non-harmonic loads). Switching ON/OFF a linear load causes significant changes in RMS current and current THD, even if the load does not produce any harmonics. TDD is an appropriate term to analyze harmonics in voltage or current waveforms, which compares root-sum-square value of the harmonic current to the maximum demand load current as represented by (\ref{TDD})~\cite{1242609}. Changes in non-harmonic loads do not cause any change in TDD as it is a normalized value.

\vspace{-0.4cm}
\begin{equation}
TDD = \frac{\sqrt[]{\sum_{n=2}^\infty I_{n RMS}^2}}{\text{Maximum Demand Load Current}} \times 100\%
\label{TDD}
\end{equation}

Flicker measures the level of small voltage fluctuations and expresses the reaction of human eye in response to a light source. IFL quantifies a sudden voltage change. Events that cause a dip in the voltage magnitude immediately interrupts the IFL data. Cumulative weighted probability of flicker perception levels at $0.1\%$, $1\%$, $3\%$, $10\%$ and $50\%$ of each 10-minute time period determine the value of $P_{st}$, which is expressed in (\ref{pst_equ})~\cite{FlickerMeasurement}.

\begin{myequation}
P_{st}=\sqrt{0.0314P_{0.1}+0.0525P_{1}+0.0657P_{3}+0.28P_{10}+0.08P_{50}}
\label{pst_equ}
\end{myequation}

where $P_{0.1}$, $P_{1}$, $P_{3}$,... denote the ratio of voltage magnitude change ($\Delta$V=$V_{max}$-$V_{min}$) to the base voltage ($V_{base}$). An average of $P_{st}$ over $2$ hours defines $P_{lt}$, which is expressed in (\ref{plt_equ})~\cite{FlickerMeasurement}.

\begin{equation}
P_{lt}=\sqrt[3]{\frac{1}{12}\sum\limits_{i=1}^{12} P_{st}^3}
\label{plt_equ}
\end{equation}

\subsection{Voltage Device Operation Analysis at the Feeders} \label{subsec:objective4,5}
Integration of PV systems on distribution feeders has some potential impacts on the operation of the legacy devices (voltage regulators, capacitor banks, reactors, load tap changers) in the network \cite{8600542,8600557}. For most traditional grid systems, utility companies often use OLTCs and regulators for voltage regulation on their distribution feeders. Switched capacitor banks are used for reactive power compensation which consequently affects the voltage profile of the feeder. More recently, the use of smart inverters for voltage regulation and optimization has become a viable option.
The major concerns with the use of VRs and OLTCs is the number of switching operations being carried out by these devices. The lifespan of VRs and OLTCs are directly impacted by the number of switching operations per day. According to~\cite{Stiles2005,Manbachi2015}, a high-quality VR is capable of making $2$ million mechanical switching ($273$ switching operations per day) without the need for maintenance over a 20-year life span.

Most commercially available OLTCs and VRs typically have a total of $32$ steps with a $0.625 \%$ change in voltage with each tap step. For VRs and LTCs, the turns ratio must satisfy the constraint expressed in (\ref{VRR})  \cite{8600542}.

\begin{equation}
Tap_{high} \geq Tap_{t} \geq Tap_{low}
\label{VRR}
\end{equation}

To better restrict the number of switching operations of the tap changers for optimal operations and simulation studies, the number of switching operations of the LTC within a time interval $t$ to $T$ should satisfy (\ref{Lim})  \cite{8600542}:

\begin{equation}
\sum _{t=1} ^ {T} N_{taps,t} \leq N_{max-tap,t}
\label{Lim}
\end{equation}.

where $N_{taps,t}$ is the number of tap changes made by the LTCs and $N_{max-tap,t}$ is the maximum allowable number of tap changes.

The impact of the eclipse on the switching of the capacitor banks can also be quantified within the period of interest. Capacitor bank switching constraints and reactive power injection are as given in (\ref{Lim2}) and (\ref{eq:6}), respectively \cite{8600542,8600557}.

\vspace{-0.7cm}
\begin{equation}
N_{\text{Daily Cap Switching}}=\sum _{t=1} ^ {T} N_{\text{Cap Switching},t} \leq N_{\text{Cap Switching}}^ {max}
\label{Lim2}
\end{equation}.

\vspace{-0.8cm}
\begin{equation}
 | Q_{cb max}^n| \geq |Q_{cb}^n|
\label{eq:6}
\end{equation}

where $N_{\text{Daily Cap Switching}}$ is the total number of switching steps by the switched capacitor in a day, $N_{\text{Cap Switching},t}$ is the number of switching operations at a time interval $t$, $N_{\text{Cap Switching}}^ {max}$ is the maximum total number of capacitor switching operations per day, $Q_{cb max}^n$ is the maximum allowable reactive power injection by a capacitor bank and $Q_{cb}^n$ is the actual reactive power injection by the capacitor bank.

The bus voltages at each node should also be within the ANSI C84.1-2011 \cite{8600557}:
\begin{equation}
\begin{split}
V_n^{max} \geq V_{n,t} \geq V_n^{min}\\
1.05 pu \geq V_{n,t} \geq 0.95 pu
\label{eq:5}
\end{split}
\end{equation}

where $V_n^{max}$ is the maximum allowable node voltage ($1.05$ pu), $V_{n,t}$ is the actual bus voltage at time $t$, and $V_n^{min}$ is the minimum bus voltage ($0.95$ pu).

\subsection{Reliability Analysis at the Management Areas} \label{subsec:objective2}
The solar eclipse is usually accompanied by multiple common weather changes such as temperature and air pressure changes, and a loss in solar irradiation. Here, the regression analysis is implemented to model the reliability metrics of the Miami and Daytona management areas under common weather conditions. Given a time period $T$, $\boldsymbol{N}=(n_1,...,n_T)$ is defined to be the daily reliability metric vector (for example, sustainable interruption), and $\boldsymbol{X}=(x_1,...,x_T)$ represents a common weather parameter vector. The relationship between $\boldsymbol{N}$ and $\boldsymbol{X}$ then can be mathematically defined as:
$\boldsymbol{N}=f(\boldsymbol{X},\boldsymbol{\beta})+\boldsymbol{\varepsilon}$, where $f(\cdot)$ denotes the regression function, typically polynomial and exponential functions  \cite{Harrell2015}; $\boldsymbol{\beta}$ represents the estimation parameter vector; and $\boldsymbol{\varepsilon}$ indicates a zero-sum white Gaussian noise. In this study, we mainly consider five main weather parameters including the average temperature $T$, the sustained wind speed $W$, the daily rain precipitation $P$, the average air pressure $A$, and the daily number of lightning strikes $L$. Taking the daily number of sustainable interruptions from January 1, 2015 to April 30, 2017 as $\boldsymbol{N}$, the relationship function between $\boldsymbol{N}$ and each weather parameter can be calculated as~\cite{wei2018hybrid}:

\begin{equation}
\begin{split}
N_{T}&=\beta^{T}_0+\beta^{T}_1{{T}}+\beta^{T}_2{{T}^2}+\beta^{T}_3{{T}^3}\\
N_{W}&=\beta^{W}_0+\beta^{W}_1\exp(\beta^{W}_2{W})+\beta^{W}_3\exp(\beta^{W}_4{W})\\
N_{P}&=\beta^{P}_0+\beta^{P}_1\exp(\beta^{P}_2{P})+\beta^{P}_3\exp(\beta^{P}_4{P})\\
N_{A}&=\beta^{A}_0+\beta^{A}_1{{A}}+\beta^{A}_2{{A}^2}+\beta^{A}_3{{A}^3}\\
N_{L}&=\beta^{L}_0+\beta^{L}_1{{L}}+\beta^{L}_2{{L}^2}+\beta^{L}_3{{L}^3}
\label{eq:regre}
\end{split}
\end{equation}

where, for $T$, $A$, and $L$, the relationship function is represented as a polynomial regression function with $3$-th degree, while, for $W$ and $P$, the relationship function is represented as the exponential regression function with $2$ terms.

Taking the common weather parameters $T$, $W$, $P$, $A$ and $L$, and their corresponding regression results $N_{T}$, $N_{W}$, $N_{P}$, $N_{A}$ and $N_{L}$ derived by (\ref{eq:regre}) as inputs, MLP is developed for forecasting the daily number of sustainable interruptions ${N}$. The MLP is a feed-forward neural network containing the input, hidden, and output layers \cite{itec,eeeic}. The proposed MLP model consists of the $10$ features mentioned earlier in the input layer ($\ell = 1$), one hidden layer with $5$ units, and an output layer, $N_{pre}$ ($\ell = 3$) with $1$ unit. All the weights $(w^{(l)}_{ji})$ between layers $\ell$ = $1$ and $\ell$ = $3$ are initialized to a set of sample values that are drawn from specific distributions instead of being randomly initialized. To achieve faster convergence and avoid saturation of activation functions during training, this paper uses Xavier uniform~\cite{xavierdist} and uniform distributions~\cite{uniformdist} to initialize the weights of MLP. To minimize the loss function (mean square error, MSE) between predicted and actual outputs during testing and training, backpropagation is used. In the input layer, the MLP neurons receive the common weather parameters and their regression results for analysis. The neurons of the output layer provide the network results for the daily number of sustainable interruptions. In the hidden layer, the MLP neurons represent the relationship between the inputs and outputs in the network. In addition, all MLP neurons are implemented with non-linear activation functions (sigmoid and hyperbolic tangent~\cite{sigmoidal}) and each MLP layer is fully connected to the next layer without the use of dropout regularization. The mathematical expression of the outputs of the MLP can be defined as~\cite{wei2018hybrid}:

\begin{equation}
N=F(b+\sum_{j=1}^{m}v_j[\sum_{i=1}^{n}{G(w_{ij}x_i+b_j)}])
\end{equation}

where $(x_1,...,x_n)$ denotes the input vector including $T$, $W$, $P$, $A$, $L$, $N_{T}$, $N_{W}$, $N_{P}$, $N_{A}$, and $N_{L}$; $N$ represents the output value; $w_{ij}$, $j=1,...,m$, is the weight of connection between the $i^{th}$ input neuron and $j^{th}$ hidden neuron; $v_j$ is the weight of connection between the $j^{th}$ hidden neuron and output neuron; $b$ and $b_j$ are the bias values of the corresponding output neuron and $j^{th}$ hidden neuron; and $F(\cdot)$ and $G(\cdot)$ are the activation functions of output and hidden neurons, respectively. An MLP architecture can contain more than one hidden layers between the input and output layers. In this study, to restrict the net capacity, one hidden layer is included in the MLP, and BP algorithm \cite{hirose1991back} is used to train the MLP architecture parameters including $w_{ij}$, $v_{ij}$, $b$, and $b_j$, where $i=1,...,n$, and $j=1,...,m$.

\vspace{-0.5cm}
\section{Case Study Results} \label{sec:results}
This section discusses the results derived from applying the formulated models to the systems of concern.

\subsection{PV System Performance} \label{subsec:objective1r}
Table \ref{tab:ppiprofiles} shows the PPI for Systems $\mathcal{A}$ and $\mathcal{B}$ during the peak of the eclipse. During the respective moments of eclipse peak (2:58 PM for System $\mathcal{A}$ and 2:45 PM for System $\mathcal{B}$), the PPI increases, which is reflected in the values recorded at the next minute (2:59 PM for System $\mathcal{A}$ and 2:46 PM for System $\mathcal{B}$). Following from (\ref{eq:ppi}), in ideal cases, the value of PPI must be close to unity because it is the ratio of observed to the expected power at a given point in time. However, the reality could be quite different, given the influence of different external factors and errors in data acquisition. These factors might cause the expected generation to exceed or fall short of the actual generation, thereby tipping the ratio to be less than unity or greater than unity, respectively. By observing the PPI values for the two systems, it can be concluded that the values of PPI are close to unity for System $\mathcal{A}$, implying that the estimation model shows good performance. However, the PPI values of System $\mathcal{B}$ exceed unity, implying the estimated values are lower than the observed values. Investigating the factors influencing the poor performance of the estimation model for this particular PV system will be part of the future work.

\begin{figure*}[!b]
  \begin{center}
  \subfloat[]{\label{fig:irrpeak}\includegraphics[width=3in, height=2in]{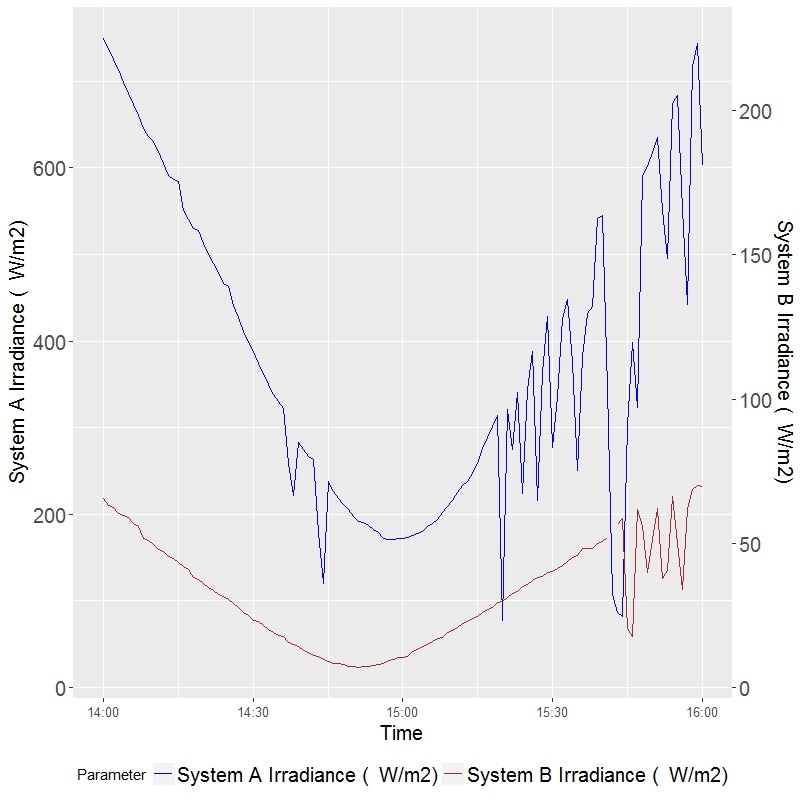}}\hspace{1cm}
  \subfloat[]{\label{fig:ambpeak}\includegraphics[width=3in, height=2in]{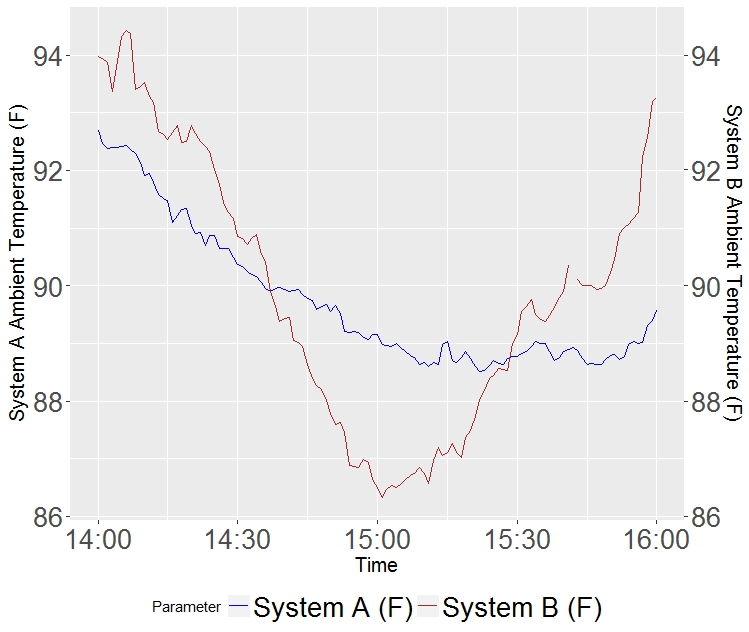}}\hspace{1cm}
  \subfloat[]{\label{fig:modpeak}\includegraphics[width=3in, height=2in]{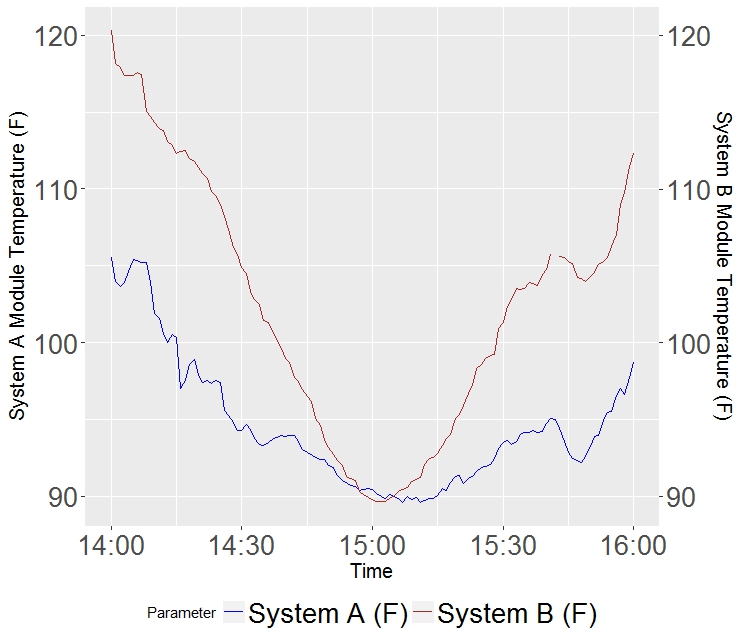}}\hspace{1cm}
   \subfloat[]{\label{fig:ppipeakpeak}\includegraphics[width=3in, height=2in]{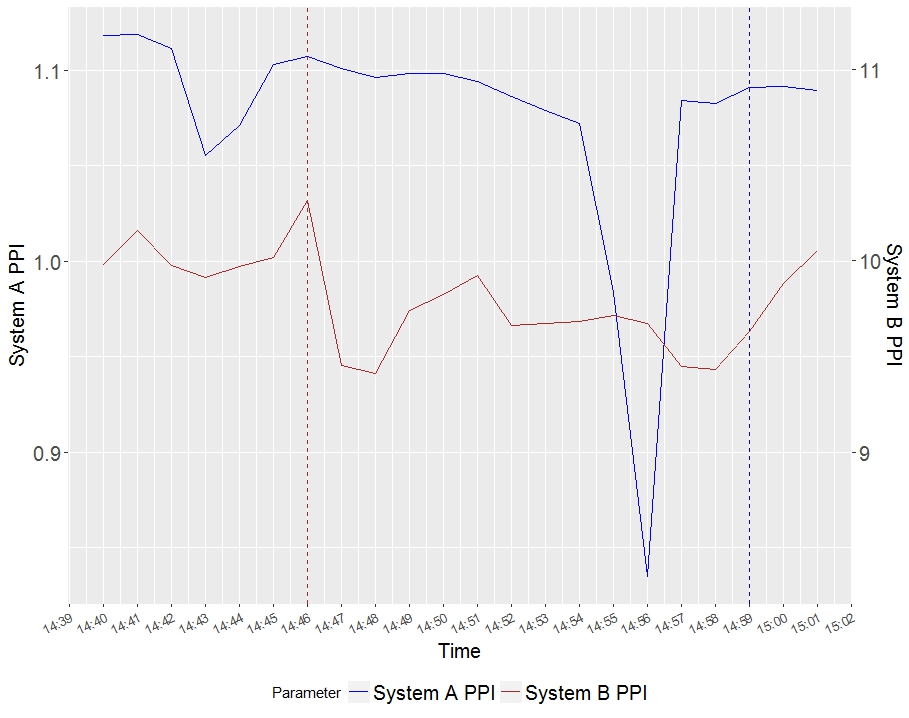}}
    \caption{\label{fig:peak} Profiles of different parameters during eclipse peak: (a) Irradiance; (b) Ambient Temperature; (c) Module Temperature; (d) PPI.}
    \vspace*{-2em}
  \end{center}
\end{figure*}

\begin{figure*}[!t]
  \begin{center}
  \subfloat[]{\label{fig:bivdur}\includegraphics[width=3.3in, height=3.75in]{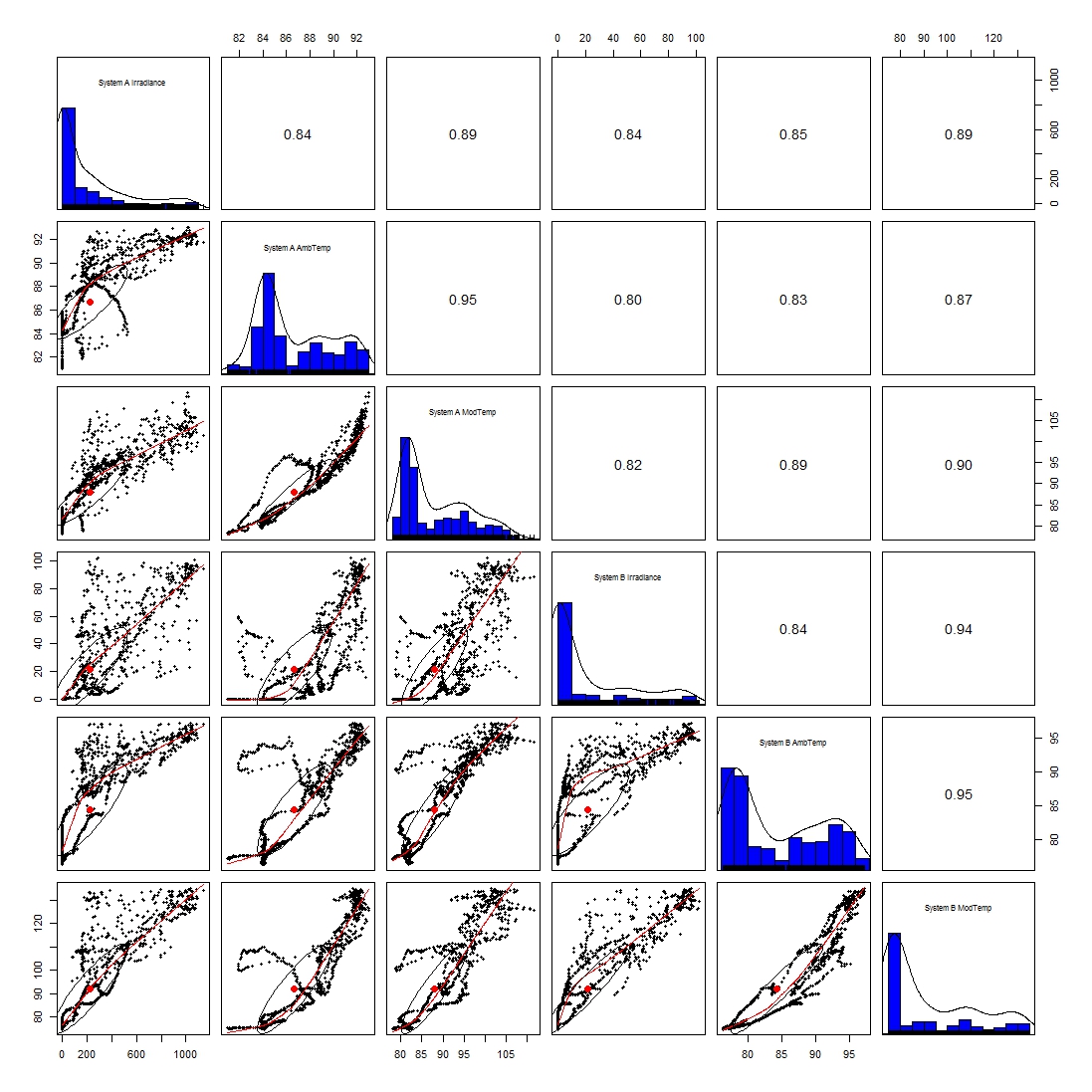}}\hspace{0.1cm}
  \subfloat[]{\label{fig:bivpeak}\includegraphics[width=3.3in, height=3.75in]{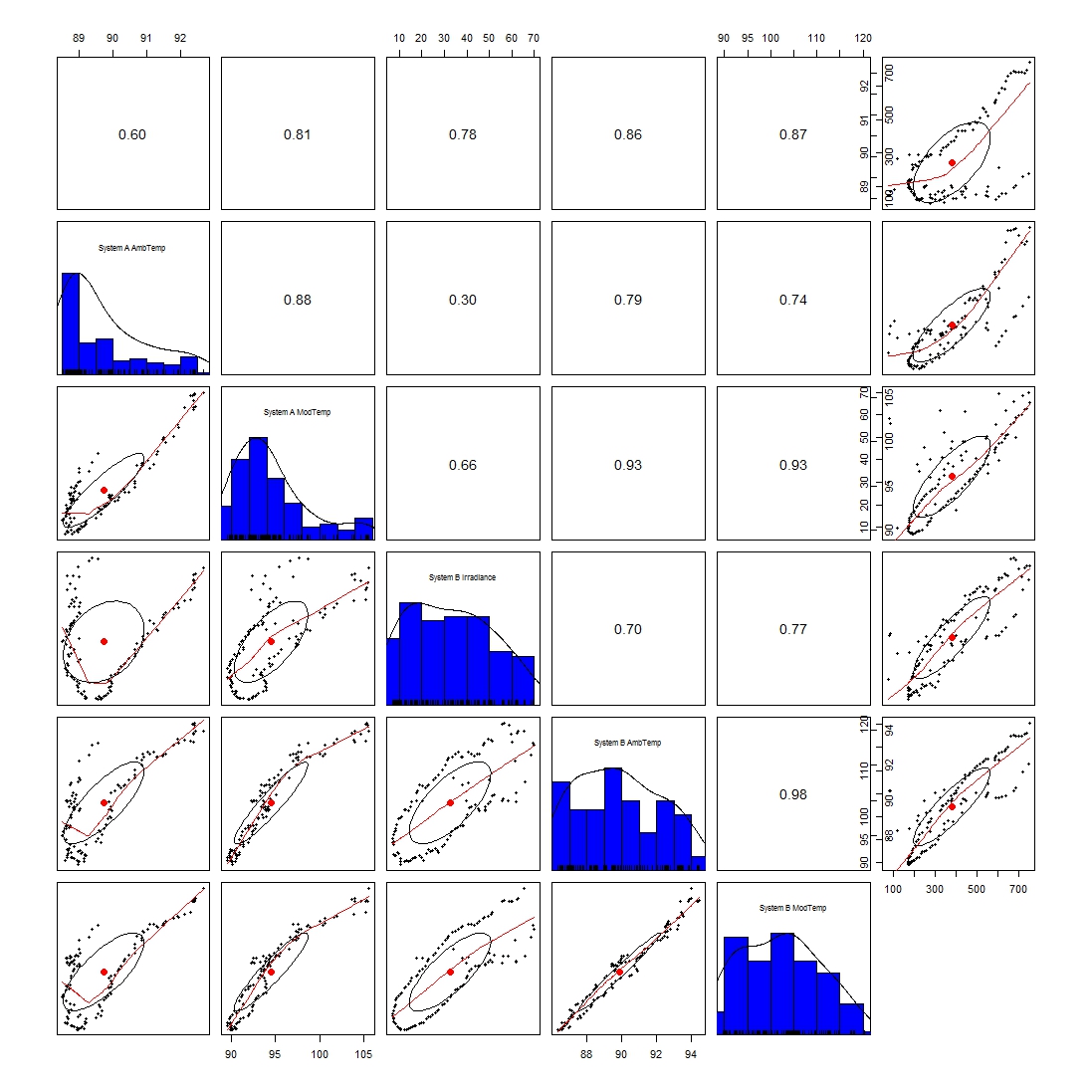}}
    \caption{\label{fig:relationship} Relationship between the different weather parameters during eclipse period and peak: (a) Bi-variate scatter plots during eclipse period; (b) Bi-variate scatter plots during eclipse peak}
    \vspace*{-2em}
  \end{center}
\end{figure*}

\begin{table}[b!]
\renewcommand{\arraystretch}{1.3}
\caption{Power Performance Index for Systems $\mathcal{A}$ and $\mathcal{B}$}
\label{tab:ppiprofiles}
\centering
\begin{tabular}{| >{\centering\arraybackslash}p{2cm}| >{\centering\arraybackslash}p{2cm}| >{\centering\arraybackslash}p{2cm}|}
\hline
\textbf{Time (PM)} & \textbf{System $\mathcal{A}$} & \textbf{System $\mathcal{B}$}\\\hline
2:44 & 1.071 & 9.975\\\hline
\textcolor{brown}{\textbf{2:45}} & 1.103 & \textcolor{brown}{\textbf{10.018}}\\\hline
\textcolor{brown}{\textbf{2:46}} & 1.107 & \textcolor{brown}{\textbf{10.319}}\\\hline
2:47 & 1.101 & 9.452\\\hline
2:57 & 1.084 & 9.676\\\hline
\textcolor{blue}{\textbf{2:58}} & \textcolor{blue}{\textbf{1.082}} & 9.447\\\hline
\textcolor{blue}{\textbf{2:59}} & \textcolor{blue}{\textbf{1.091}} & 9.629\\\hline
3:00 & 1.091 & 9.885\\\hline
\end{tabular}
\end{table}

The variations of different measured parameters during the peak for the two systems is shown in Fig. \ref{fig:peak}. While Figs. \ref{fig:modpeak}, \ref{fig:ambpeak} and \ref{fig:ppipeak} show the variations of the module temperature, ambient temperature and PPI from 2 PM through 4 PM, Fig. \ref{fig:ppipeakpeak} shows the variation of PPI during the eclipse peak minute-by-minute. The module temperature and PPI for systems $\mathcal{A}$ and $\mathcal{B}$ show a strong visual correlation while the ambient temperature curves show a steeper dip for System $\mathcal{B}$ than for System $\mathcal{A}$. These visual correlations imply a strong positive relationship during the eclipse between such weather parameters of the two geographically separate PV systems. This correlation, although not an indication of dependency, is a marker of how the two PV systems can be aggregated or utilized in combination during the eclipse to address associated loads and other aspects in the future high penetration scenarios.

To further explore the relationship between irradiance, ambient temperature and module temperature for Systems $\mathcal{A}$ and $\mathcal{B}$ (a total of 6 variables), bi-variate scatter plots were done for both during the eclipse duration between 2:00 and 3:00 PM as well as during eclipse peak between 2:40 and 3:00 PM, which are illustrated in Figs. \ref{fig:bivdur} and \ref{fig:bivpeak}, respectively. The plots show a 6x6 matrix with the PDF of the 6 variables along the primary diagonal, scatter plots with model fitting below the diagonal, and the Pearson correlation coefficient above the diagonal. The model fitting shows that all of the relations between parameters within and between the systems are linear. However, this relationship changes during the eclipse peak where the ambient temperature of System $\mathcal{A}$ has a curvilinear relationship with other parameters of both $\mathcal{A}$ and $\mathcal{B}$. Overall, the correlation coefficients of the parameters of the two systems show a drop during the eclipse peak, suggesting the two systems have a less strong positive relationship during this period and hence could be used in aggregation-related studies. For example, the power from System $\mathcal{A}$ could be used to meet the deficit observed at System $\mathcal{B}$ provided the cost of power transfer is less than other alternatives.

\subsection{Power Quality Analysis at the POI} \label{subsec:objective3r}
Impacts of the eclipse on power quality metrics at the POI of System $\mathcal{A}$ are analyzed in the following subsections. Observations on the average RMS voltage, harmonics (both voltage and current), TDD, and flicker are also discussed.

\begin{figure*}[!b]
  \begin{center}
  \subfloat[]{\label{AVG_Voltage_MaxCoverage}\includegraphics[width=3in, height=1.25in]{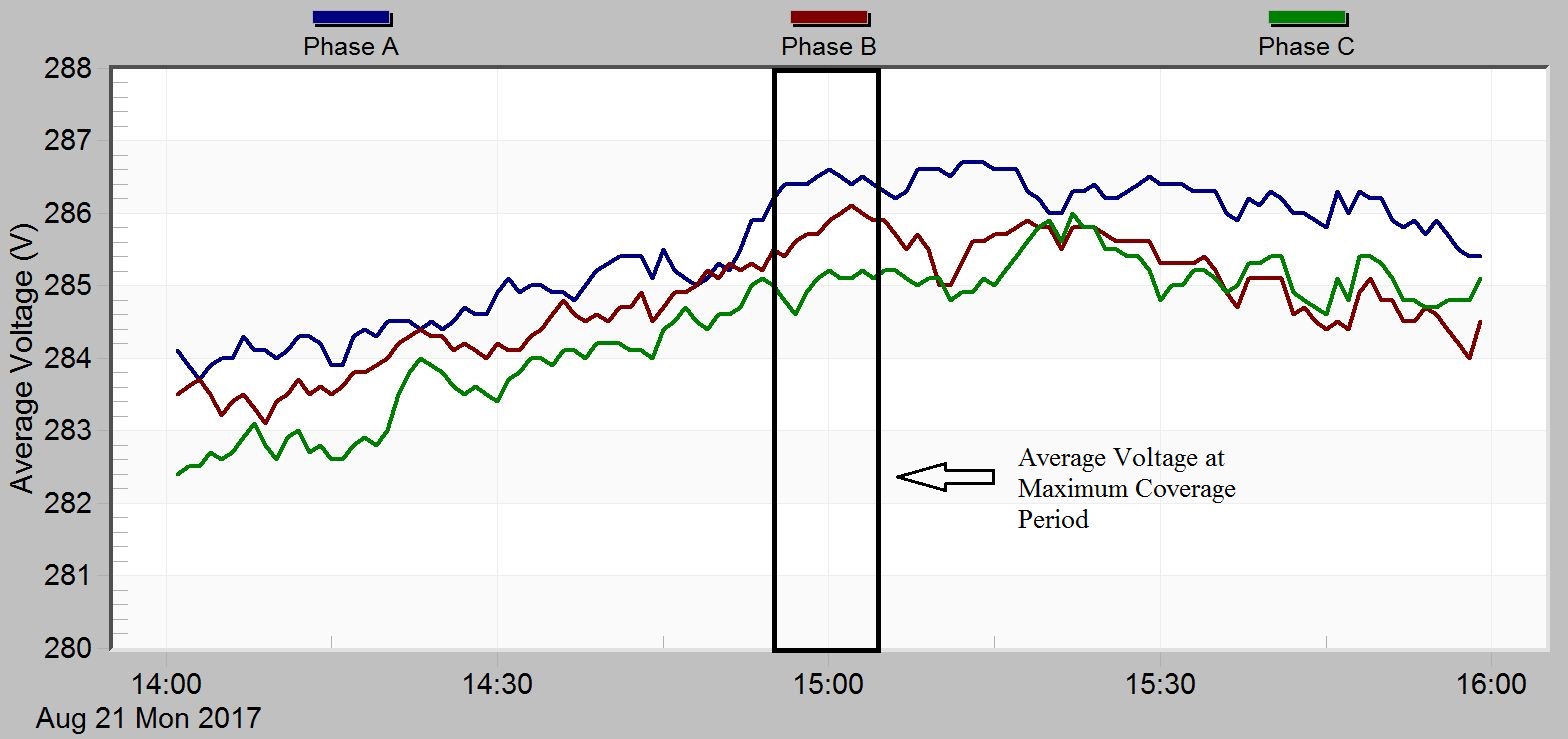}}\hspace{1cm}
  \subfloat[]{\label{Voltage_THD}\includegraphics[width=3in, height=1.25in]{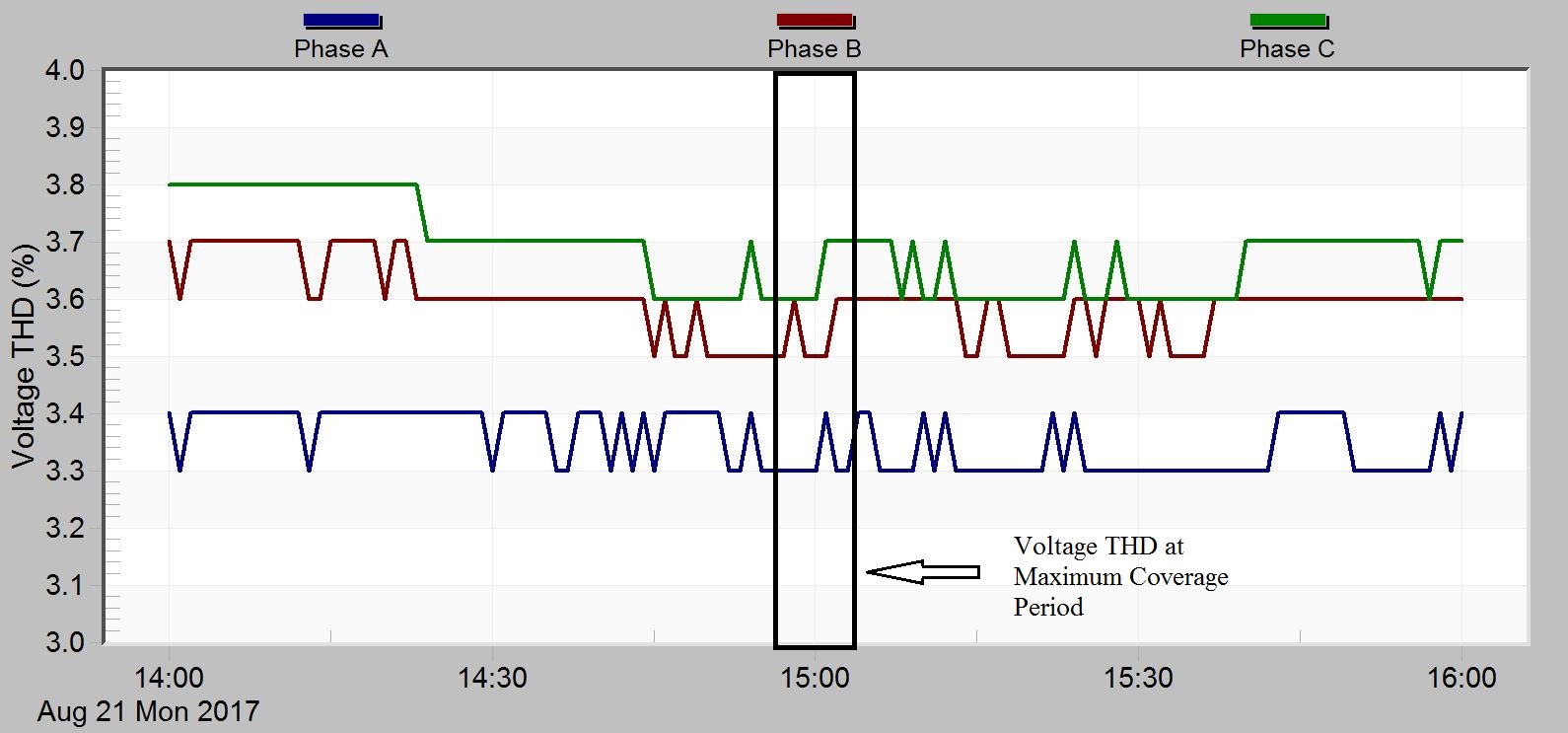}}\hspace{1cm}
  \subfloat[]{\label{Current_THD}\includegraphics[width=3in, height=1.25in]{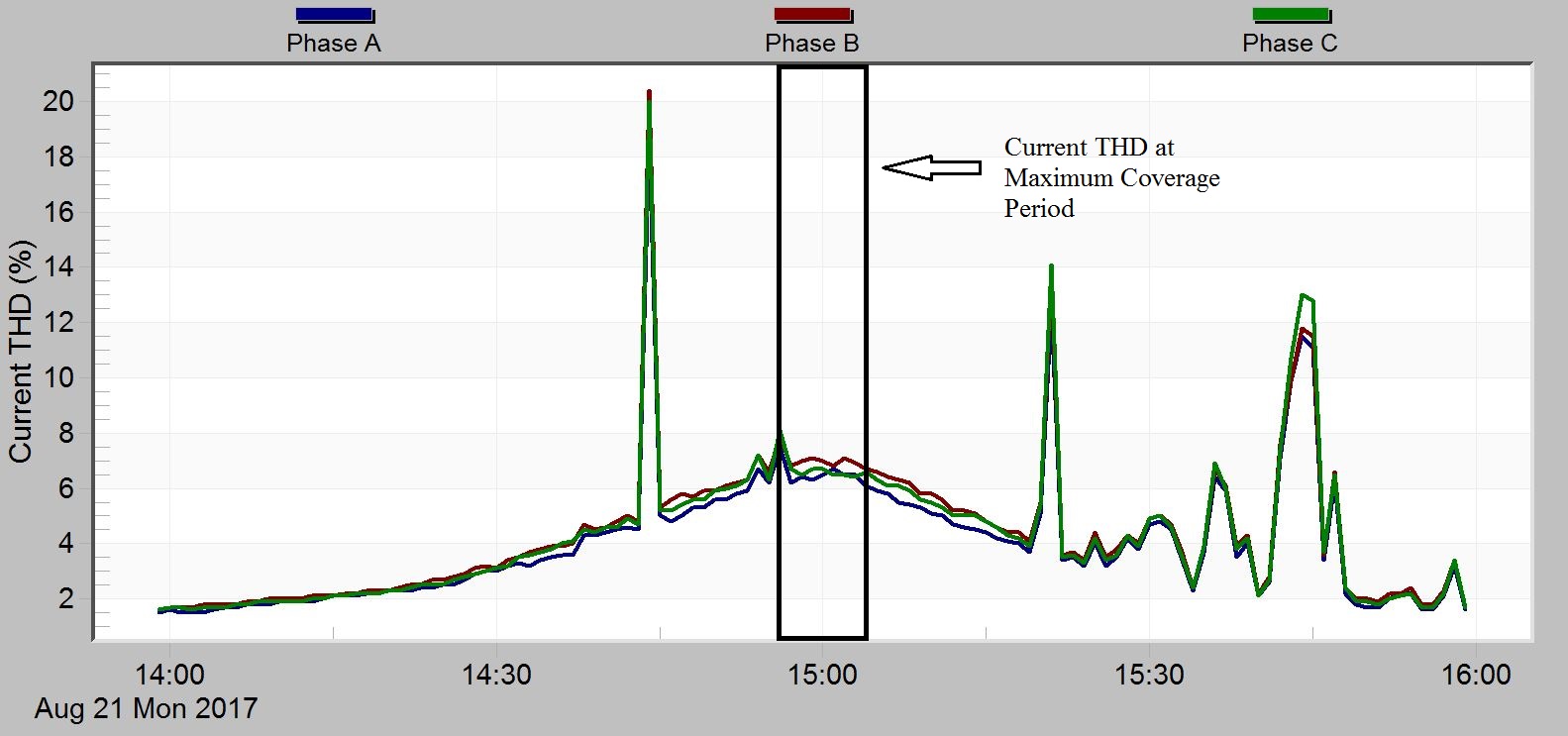}}\hspace{1cm}
   \subfloat[]
   {\label{IFL}\includegraphics[width=3in, height=1.25in]{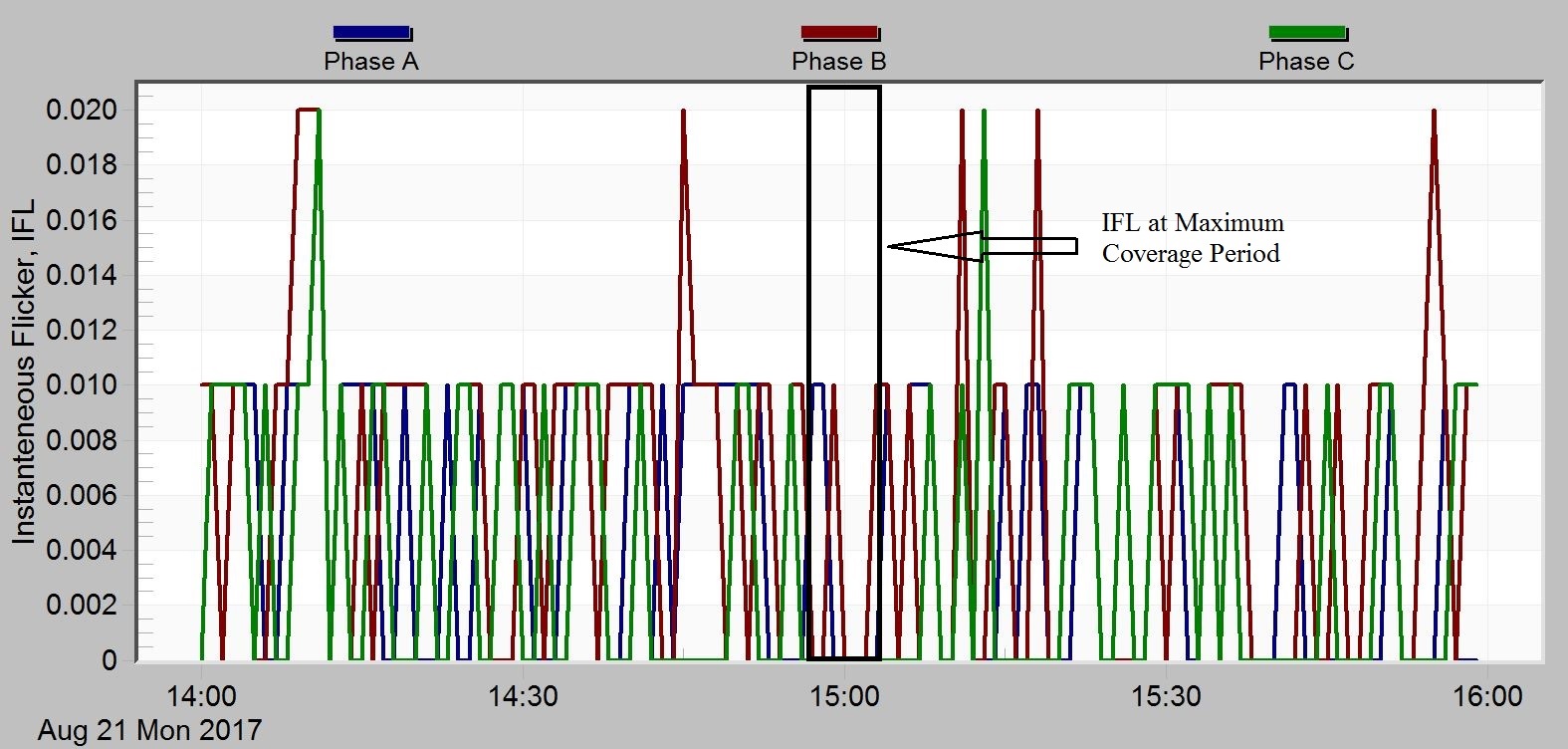}}
    \caption{\label{} Profiles of different power quality metrics at POI during solar eclipse period of grid-tied System $\mathcal{A}$: (a) Average RMS Voltage; (b) Voltage THD; (c) Current THD; (d) IFL.}
    \vspace*{-2em}
  \end{center}
\end{figure*}

\begin{figure*}[!b]
  \begin{center}
   \subfloat[]
   {\label{Pst}\includegraphics[width=3in, height=1.25in]{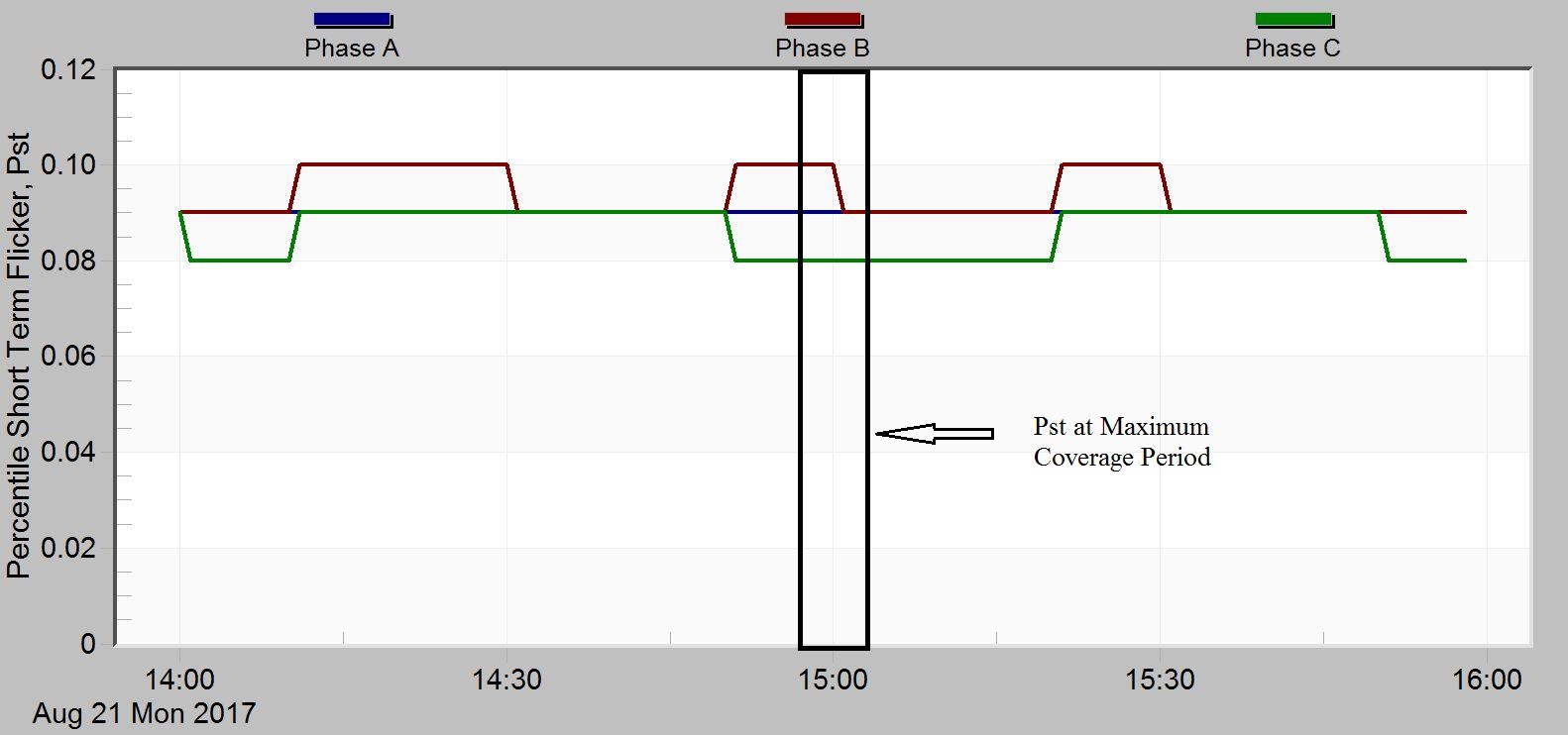}} \hspace{1cm}
   \subfloat[]
   {\label{Plt}\includegraphics[width=3in, height=1.25in]{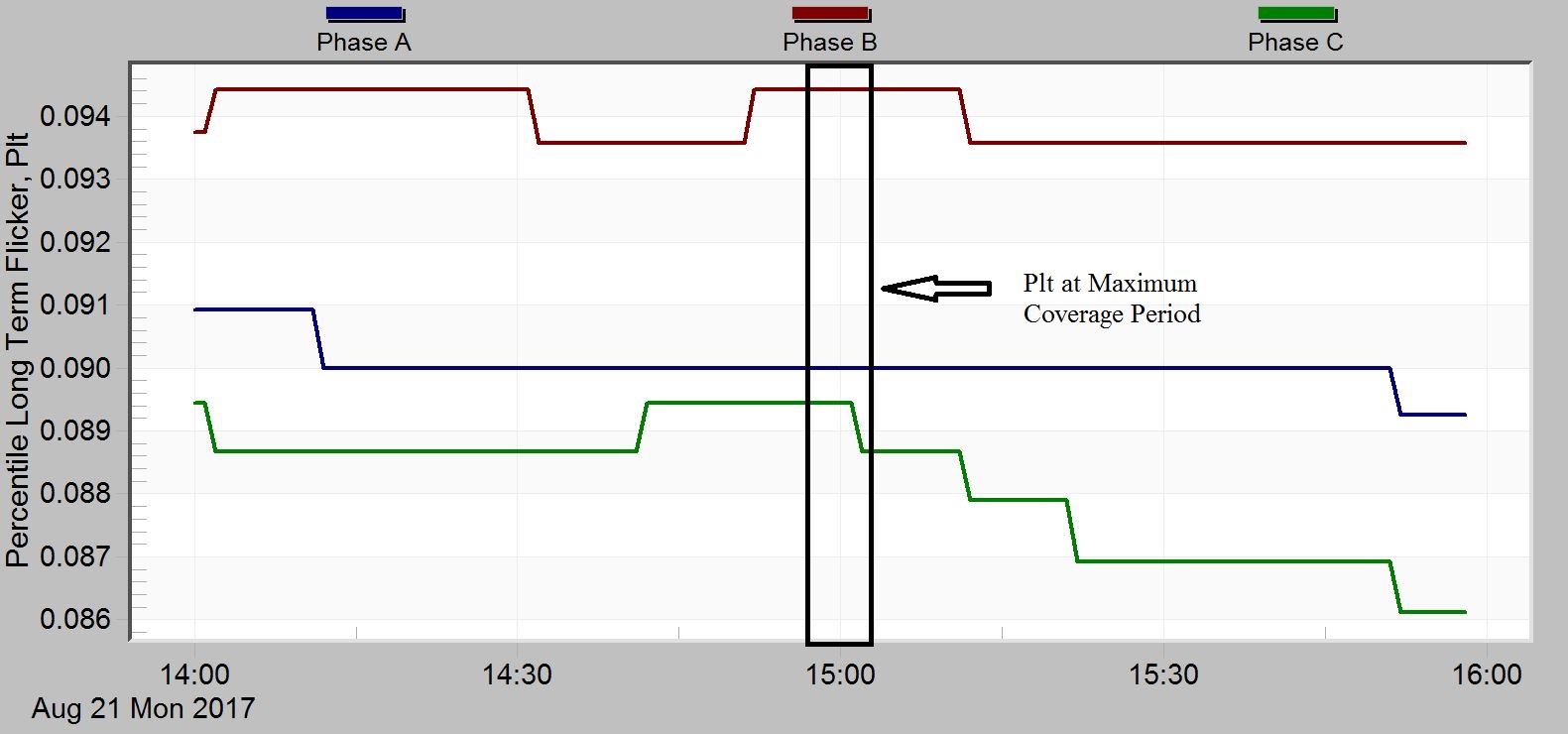}}
    \caption{\label{} Profiles of different power quality metrics at POI during solar eclipse period of grid-tied System $\mathcal{A}$: (a) $P_{st}$; (b) $P_{lt}$.}
    \vspace*{-2em}
  \end{center}
\end{figure*}

\subsubsection{Average Voltage} \label{subsec:avgvolt}
PV generation affects the feeder voltage where it is connected to. According to IEEE Standard 1547-2018, the feeder voltage at a low voltage distribution network can vary within $\pm5\%$ of the base voltage. Therefore, the minimum and maximum allowable ranges of average RMS voltage at the POI are $265 V$ and $292 V$, respectively. The average RMS voltage at the POI during the eclipse is illustrated in Fig. \ref{AVG_Voltage_MaxCoverage}. The average voltage was expected to drop at maximum coverage due to reduced generation but an opposing trend of voltage change is observed. System $\mathcal{A}$ has a marginal impact on voltage change at the POI during the eclipse.

\begin{figure}[t!]
    \centering
\includegraphics[width=8cm,height =5cm ]{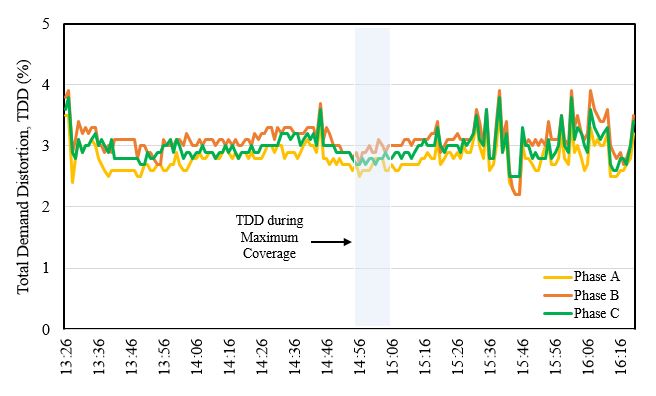}
    \caption{Total Demand Distortion during Solar Eclipse Period.}
    \label{TDD}
\end{figure}

\subsubsection{Harmonics}
Current THD is expected to increase rapidly during the eclipse peak due to low solar irradiance and PV generation, because THD is inversely proportional to the fundamental RMS current value of PV inverters ~\cite{6939147,pvsys1}. On the other hand, voltage THD should vary very little as the voltage at the POI is allowed to change in small ranges. The voltage and current THDs during the eclipse are shown in Figs. \ref{Voltage_THD} and \ref{Current_THD}, respectively. The measured voltage THD varies between $3.4\%$ and $4\%$, which is within the IEEE Standard 1547-2018 for voltage THD ($5\%$). A large range of current THD value (from $2\%$ to $35\%$) was observed during the eclipse. Several sharp spikes were identified in Fig. \ref{Current_THD} because current distortion is very sensitive to the frequent changes of incident solar radiation (caused by cloud movements) \cite{patsalides2007effect}. To investigate the current distortion impact of solar eclipse, the TDD is plotted as shown in Fig. \ref{TDD}. The TDD curve is relatively flat during the eclipse except some minor fluctuations. If the PV System $\mathcal{A}$ provides a larger percentage of the total demand current, the TDD effect would become prominent.

\subsubsection{Flicker}
Figures \ref{IFL}, \ref{Pst}, and \ref{Plt} show three flicker indices during the eclipse that express small voltage variations at the POI. Both $P_{st}$ and $P_{lt}$ are within $1.0$ and $0.8$, respectively and are within the IEEE Standard 1453-2015 voltage fluctuation limit. The variations observed in these three figures seem very small but the scenario could be different at high penetration during the solar eclipse.

The values of different power quality parameters at the POI of System $\mathcal{A}$ during the start, end, and peak of the eclipse, along with the corresponding IEEE standards limits, are presented in Table \ref{Power Quality}. It clearly indicates that the quality of power at the POI of System $\mathcal{A}$ during the eclipse was within allowable IEEE standard at the existing penetration level.

\begin{table}[t!]
\tiny
\caption{Power Quality Metrics Statistics during Solar Eclipse of 21 August, 2017}
\label{Power Quality}
\begin{center}
\begin{tabular}{|c|c|c|c|c|c|}
\hline
 \begin{tabular}{c} \textbf{Power} \\ \textbf{Quality} \\ \textbf{Metrics}\end{tabular} & \textbf{Phase} & \begin{tabular}{c}\textbf{Start Time} \\ \textbf{(1:26 PM)} \end{tabular} & \begin{tabular}{c}\textbf{Maximum} \\\textbf{Coverage} \\ \textbf{(2:58 PM)} \end{tabular} & \begin{tabular}{c}\textbf{End Time} \\\textbf{(4:20 PM)}\end{tabular} & \begin{tabular}{c} \textbf{IEEE}\\ \textbf{Standard}\end{tabular}\\
\hline
\multirow{3}{*}{\begin{tabular}{c}Average \\ RMS \\ Voltage \end{tabular}} & A & 283.4 & 286.8 & 284.8 & $265\sim292$ \\ & B & 283.1 & 286.0 & 284.4 & $265\sim292$\\ & C & 282.9 &  285.3 & 284.3 & $265\sim292$ \\
\hline
\multirow{3}{*}{\begin{tabular}{c}Voltage \\ THD (\%) \end{tabular}} & A & 3.5 & 3.4 & 3.5 & 5.0 \\ & B & 3.7 & 3.6 & 3.7 & 5.0 \\ & C & 3.8 &  3.7 & 3.8 & 5.0\\
\hline
\multirow{3}{*}{\begin{tabular}{c}Current \\ THD (\%) \end{tabular}} & A & 4.8 & 6.7 & 5.6 & - \\ & B & 5.1 & 7.5 & 5.7 & - \\ & C & 5.1 &  7.0 & 5.7 & - \\
\hline
\multirow{3}{*}{\begin{tabular}{c}Total \\ Demand \\ Distortion\end{tabular}} & A & 3.5 & 2.6 & 3.2 & 5.0 \\ & B & 3.8 & 2.9 & 3.5 & 5.0\\ & C & 3.6 &  2.7 & 3.4 & 5.0 \\
\hline
\multirow{3}{*}{\begin{tabular}{c}IFL\end{tabular}} & A & 0.01 & 0.0 & 0.0 & - \\ & B & 0.01 & 0.0 & 0.01 & -\\ & C & 0.0 & 0.0 & 0.01 & - \\
\hline
\multirow{3}{*}{\begin{tabular}{c}$P_{st}$\end{tabular}} & A & 0.09 & 0.09 & 0.09 & 1.0 \\ & B & 0.09 & 0.1 & 0.09 & 1.0\\ & C & 0.08 &  0.08 & 0.08 & 1.0 \\
\hline
\multirow{3}{*}{\begin{tabular}{c}$P_{lt}$\end{tabular}} & A & 0.091 & 0.09 & 0.089 & 0.8 \\ & B & 0.0945 & 0.0945 & 0.0935 & 0.8\\ & C & 0.0885 &  0.0895 & 0.086 & 0.8 \\
\hline
\end{tabular}
\end{center}
\end{table}

\subsection{Voltage Regulating Devices Operation and Analysis} \label{subsec:objective4,5r}
The sharp ramps in irradiance and, consequently, the generation of the PV systems during an eclipse have some impacts on the voltage profile of the network, the operations of the legacy voltage control devices (VRs and OLTCs), and losses in the network. The severity of these impacts depend, among others, on the location of the POI of the systems, the penetration level of these systems, the control setting of the voltage regulation devices, the level of ramping of the power generation from the systems, and the load profile of the network.

\subsubsection{Voltage Regulator and LTC Operations}
The dynamic changes in the voltage profile of the network due the ramp in power output of the PV systems causes the LTCs and VRs to tap change in other to control the voltage within the ANSI C84.1-2011 voltage standard. Figure \ref{fig:feeder} shows the various tap changing done by the VRs and the substation transformer LTC. For the feeders under study, substation transformer did not tap at all. Obviously, this is due to the relatively far distance between the PVs and the substation, and the presence of other voltage control devices close to the PVs to affect some voltage regulation. A tap change on the substation transformer is always the last resort for voltage control, because a change in tap affects the whole feeder voltage downstream the network.
Of the three voltage regulators downstream the feeder, Vreg 3 tapped the most (7 times on phase A, 5 times on phase B and none on phase C) during the eclipse. This was due to its proximity to System $\mathcal{A}$ (coupled with its relatively large size compared to System $\mathcal{B}$) and Cap 0 . Vreg 3 tapped 3 times on phase A, no tap change on phase B and phase C during the simulated event. Vreg 4 did not tap change at all during the simulated event. This could be partly due to the voltage control action done by Cap 3.

\begin{figure*}[t!]
    \centering
\includegraphics[width=15cm,height = 12cm ]{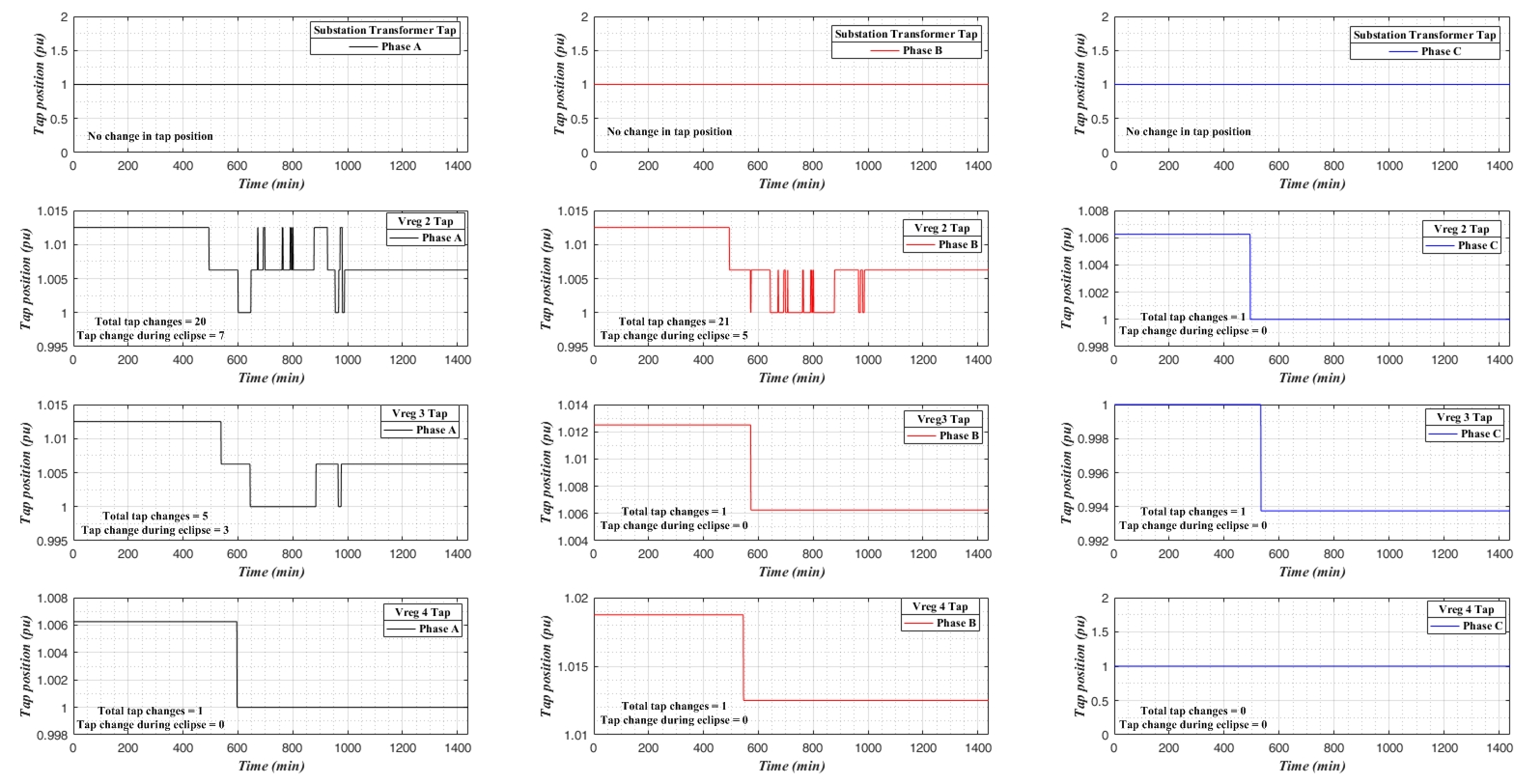}
    \caption{LTC and Voltage regulator tap changing during the eclipse}.
    \label{fig:feeder}
\end{figure*}

\begin{figure*}[t!]
    \centering
\includegraphics[width=15cm,height = 10cm ]{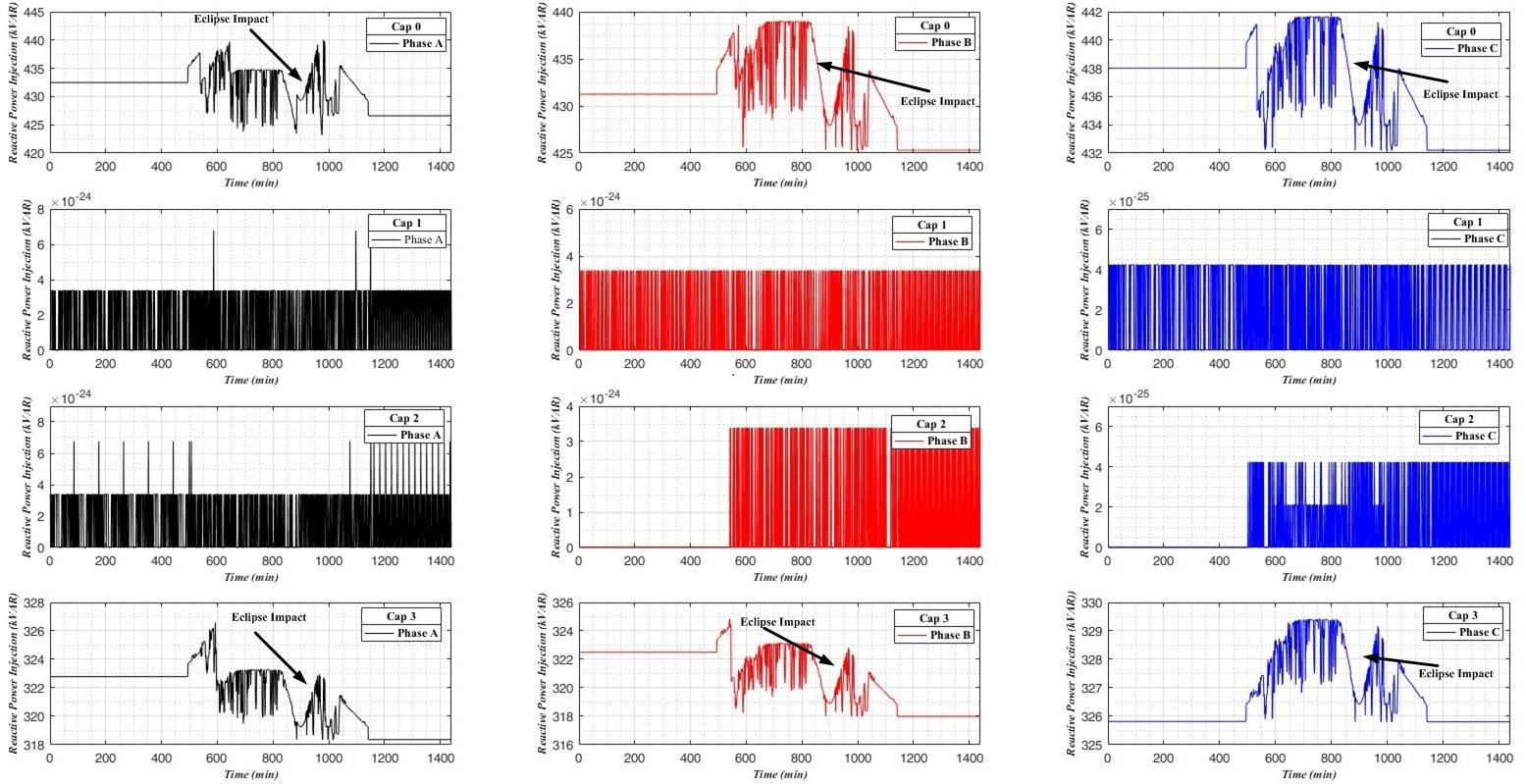}
    \caption{Capacitor bank reactive power injection during the eclipse}.
    \label{fig:feeder}
\end{figure*}

\begin{figure*}[t!]
    \centering
\includegraphics[width=16cm,height = 6cm ]{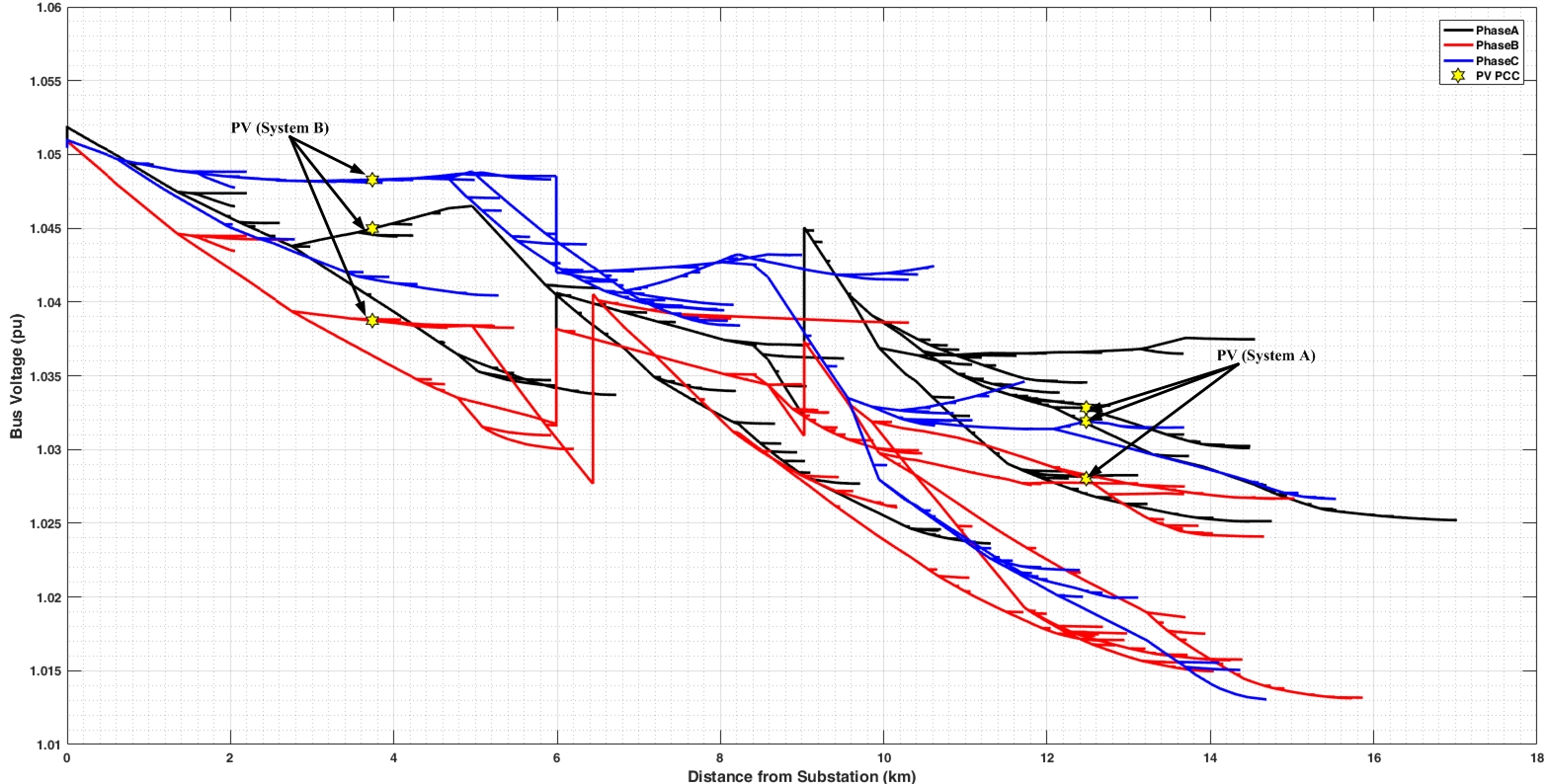}
    \caption{Feeder voltage profile at the peak of eclipse on PV (System A)}.
    \label{fig:feeder Voltage A}
\end{figure*}

\begin{figure*}[t!]
    \centering
\includegraphics[width=16cm,height = 6cm ]{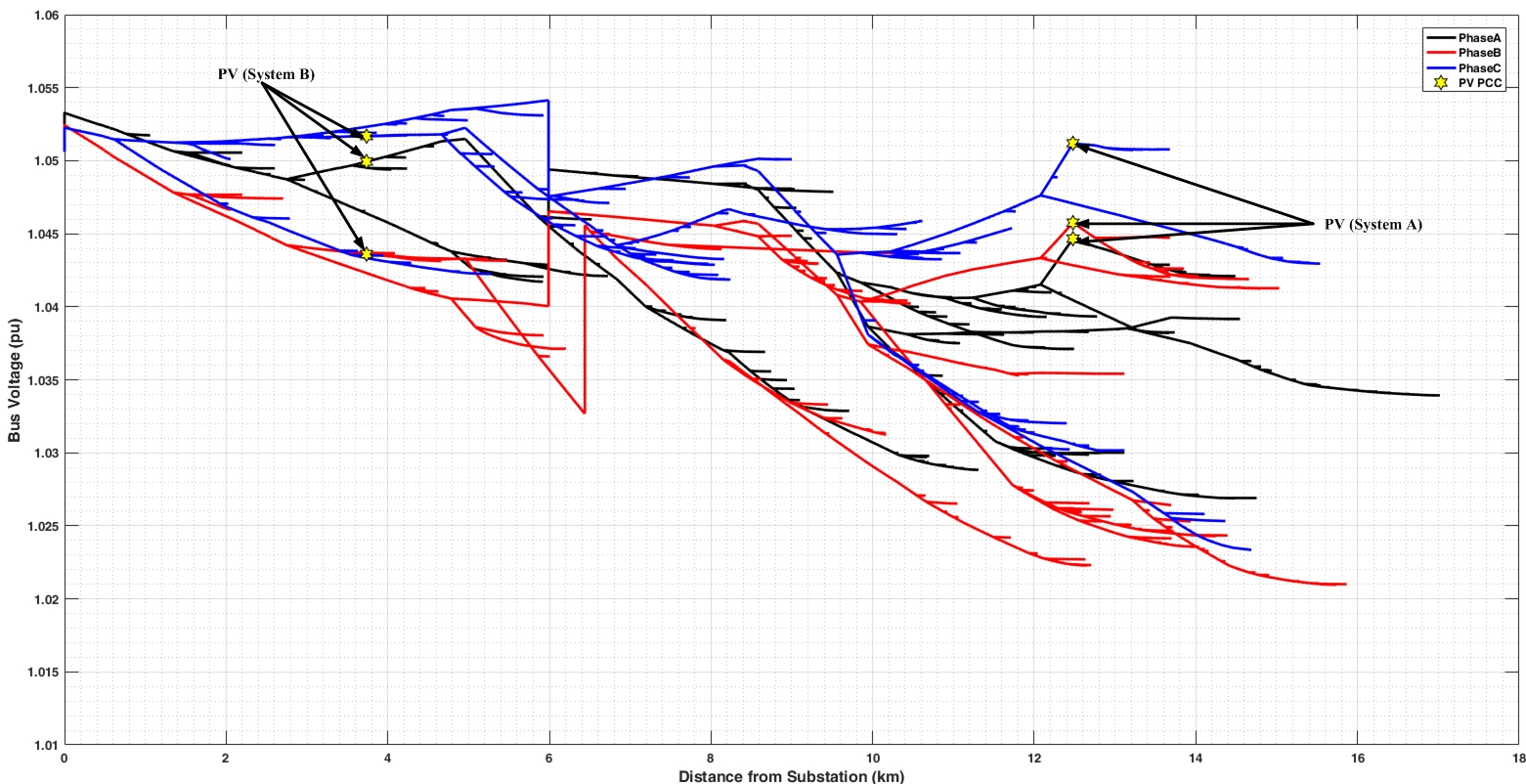}
    \caption{Feeder voltage profile at the peak of eclipse on PV (System B)}.
    \label{fig:feeder Voltage B}
\end{figure*}

\begin{figure}[t!]
    \centering
\includegraphics[width=8cm,height = 5cm ]{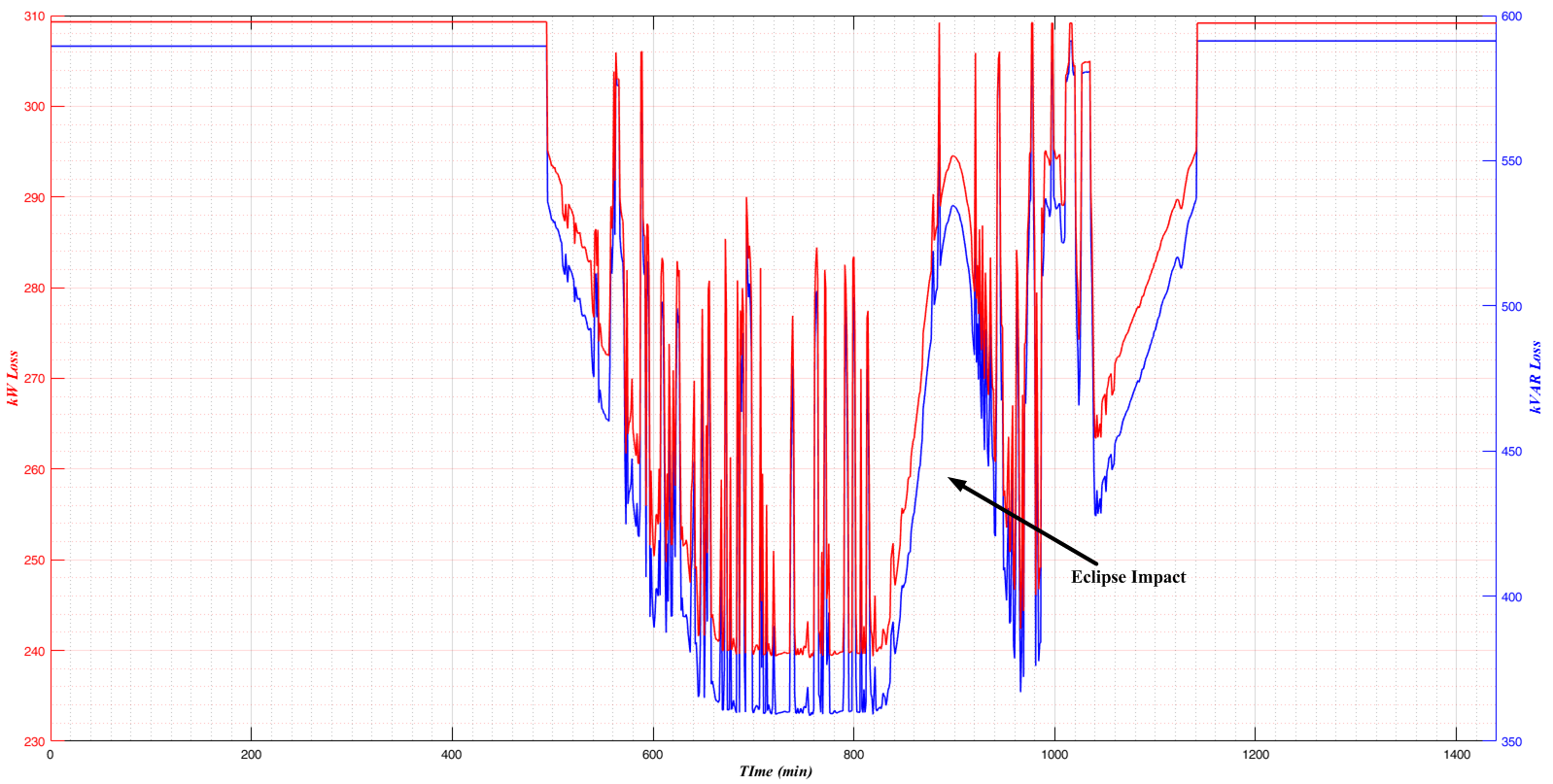}
    \caption{Active and Reactive system losses during the eclipse}.
    \label{fig:feeder Loss}
\end{figure}

\subsubsection{Capacitors Operation}
During the simulated eclipse scenario, capacitors Cap 0 and Cap 3 injected reactive power into the network. Capacitors Cap 1 and 2 injected approximately zero amount of reactive power. The proximity of Cap 1 and Cap 2 to the substation and the rigidity of the grid voltage upstream the feeder could be possibly explain this negligible injection of reactive power by  Caps 1 and 2. The capacitor control settings of Caps 1 and 2 prevented them also from injecting power due to the relatively small variation in voltage before and during the eclipse. The downstream locations of Cap 3 and 0 coupled with their proximities to Vreg 2 and 4 and the PV systems could be responsible for their reactive power injections. The reactive power injections on phase B and C of Cap 0 is higher than that of phase A. This would explain  why there was a constant tapping down on phase B and tapping up of phase A of Vreg 2.
It is interesting to note the profile of the reactive power injection by Caps 0 and 3. During the eclipse time frame, the ramps (loss in power generation from the PV systems) consequently led to a reduction in the reactive power injection by the capacitors. The constant reactive power injection as seen for the first 500 minutes of  Caps 0 and 3 was due to the relatively constant voltage of the feeder prior to power generation by the PVs.

\subsubsection{Voltage Profile}
The feeder voltages (on the buses) plotted against their distances from the substation is as shown in Figs. \ref{fig:feeder Voltage A} and \ref{fig:feeder Voltage B}. PV Systems $\mathcal{B}$ and $\mathcal{A}$ are approximately $3.75$ km and $12.5$ km from the substation, respectively. The two plots (Figs. \ref{fig:feeder Voltage A} and \ref{fig:feeder Voltage B}) shows that the voltage profile at the instant when the peak of the eclipse occurred at both PV locations respectively. At the peak of the eclipse on PV System $\mathcal{A}$, the ramp down apparently had no voltage impact on the voltage profile, since as expected the ramp down in  generation could only lead to a decrease in voltage at the point of interconnection. Also at this instant at PV System $\mathcal{B}$, its proximity to the substation makes the voltage at the point of interconnection much more rigid, reducing the possibility of a fluctuation in voltage as a result of the eclipse. Also, from Fig. \ref{fig:feeder Voltage B}, the peak of the eclipse on PV System $\mathcal{B}$ did not cause any significant impact on the voltage profile on the feeder. At this instant, the power on PV System $\mathcal{A}$ tend to ramp up which led to a sharp increase in voltage at the POI downstream of the feeder. The penetration level of these systems obviously did not produce any significant impact on the voltage profile during the eclipse event. With higher levels of penetration, these impacts could become severe.

\subsubsection{Network Losses}
The time series plot of the overall system losses (transformer losses plus the line losses)is shown in Fig. \ref{fig:feeder Loss}. Locating PVs downstream a feeder usually reduces the overall system losses by improving the voltage profile downstream. This is quite obvious from the plot. The losses is the network starts reducing at the instant of aggregated PV generation from both PV plants. The effect of the eclipse is also seen from the loss plot. During the eclipse, the short-fall in generation from the PV led to an increase in the overall system losses during the eclipse.

\subsection{Reliability Analysis at the Management Areas} \label{subsec:objective2r}

To evaluate the relationship between system reliability at the management areas and common weather parameters, the daily number of sustainable interruptions is collected from the Miami management area ranging from Jan. 1st 2015 to Apr. 30th 2017. The daily number of lightning strikes $L$ is collected from the local utility's weather station centrally located at the Miami management area. The other four common weather parameters including $L$, $W$, $P$, and $A$ are hourly sampled from the Miami international airport. After data collection, both the reliability and common weather data are integrated as inputs to train the regression models defined in (\ref{eq:regre}).

\begin{figure} [tbp] \centering
\includegraphics[width=8.5cm,height=6cm]{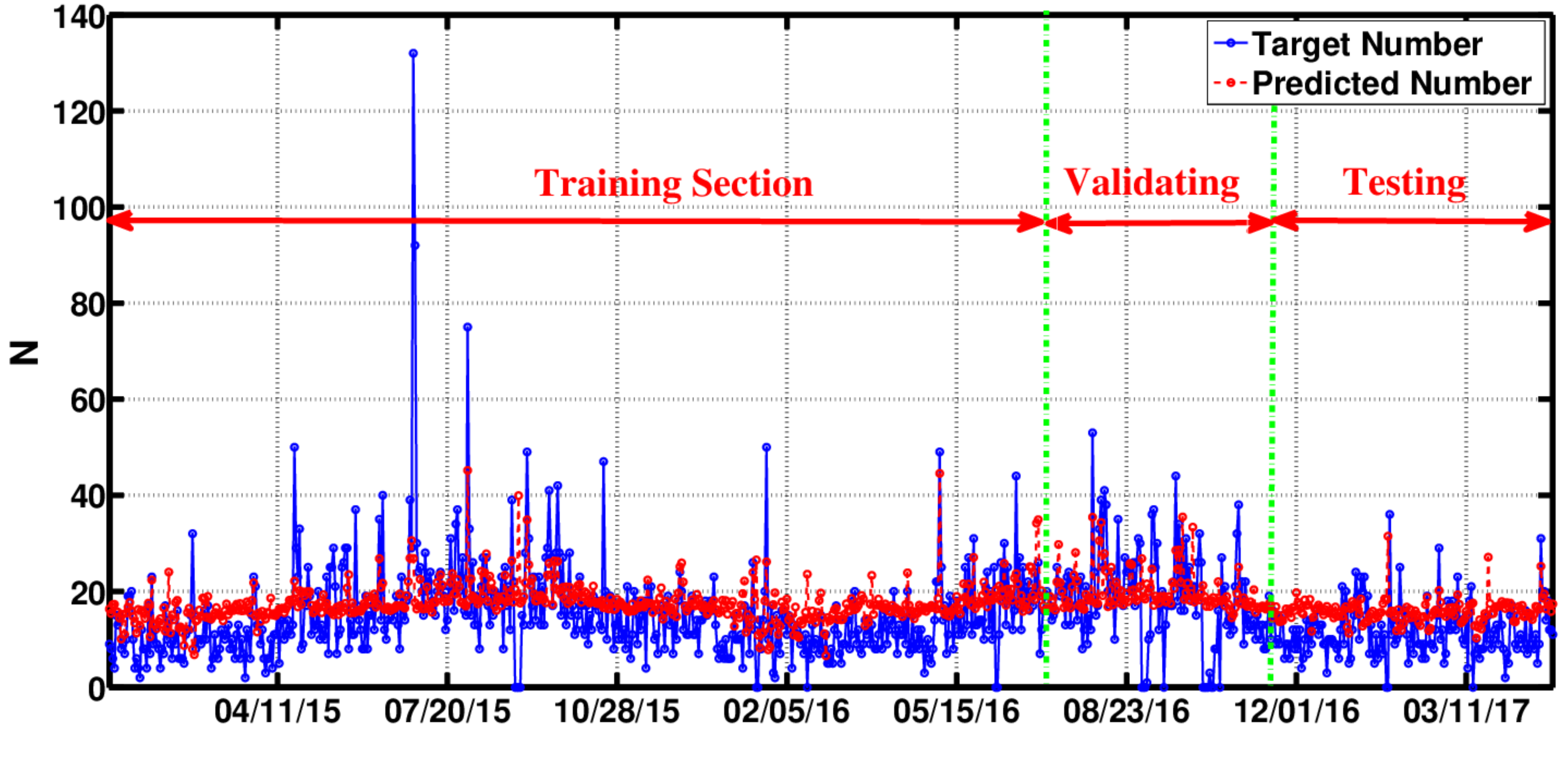}
\caption{Comparison between the target numbers of sustainable interruptions and the predicted numbers derived by the proposed MLP forecasting model.}
\label{fig:Neural}
\end{figure}

In the proposed MLP forecasting model, the input layer contains five types of common weather parameters and corresponding daily numbers of sustainable interruptions derived by their regression models. The MLP hidden layer is set with twenty neurons, and the output layer involves the target number of sustainable interruptions. The BP algorithm is introduced to train, validate, and test the proposed MLP forecasting model, where 70\% of the collected data is used for training, 15\% of the data is implemented for validating, and the remaining 15\% is applied for testing. For all training, validating, and testing sections, Fig. \ref{fig:Neural} compares the target numbers of sustainable interruptions with the predicted numbers derived by the proposed MLP forecasting model~\cite{Hybrid1}. In this figure, we can find that, the proposed MLP forecasting model provides an acceptable result in comparison with the actual numbers of sustainable interruptions in most time periods. In particular, the proposed MLP forecasting model derives a mean-squared error of $315.4$.

\begin{figure}[tbp]
\centering
\includegraphics[width=8.5cm,height=5.5cm]{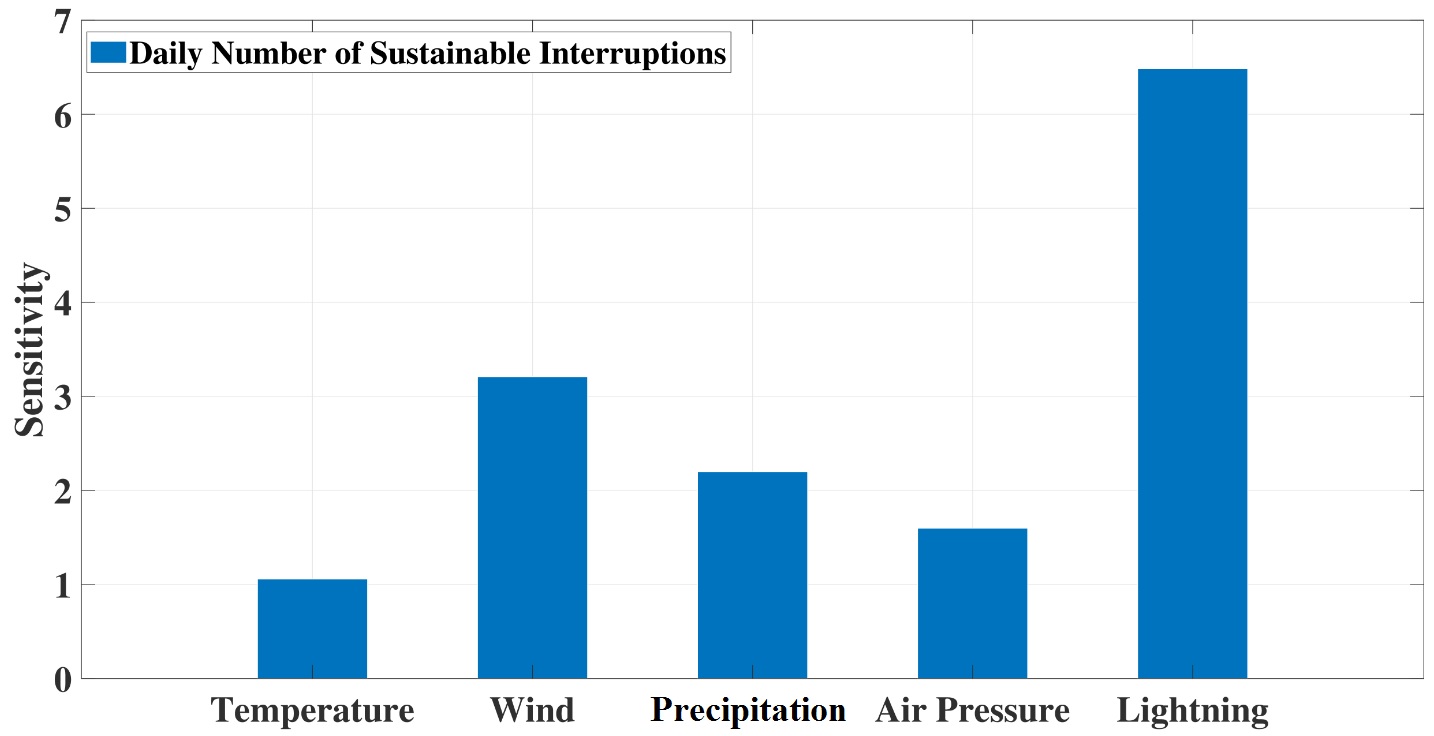}
\caption{Sensitivity analysis of daily number of sustainable interruptions with each weather parameter.}
\label{fig:Sens}
 \vspace*{-2em}
\end{figure}

In addition, the sensitivity analysis is used to evaluate the effect of each common weather parameter on the daily number of sustainable interruptions, respectively. The sensitivity value is calculated by the first-order derivative of MLP function with respect to the network parameters. Fig. \ref{fig:Sens} presents the sensitivity of each weather parameter response to the daily number of sustainable interruptions, respectively. In this figure, we can find that lightning strike $L$ is the most important weather parameter that has an influence on the daily number of sustainable interruptions, while the average temperature $T$ has the least impact on the daily number of sustainable interruptions. This phenomenon can be explained that the most numbers of sustainable interruptions happen ranging from June to September during one year, which is the raining season for the Florida and lightning strikes happen most frequently. Since the average temperature of Florida almost keeps between $80{^{o}F}$ and $95{^{o}F}$ for the most months during the year, the temperature change has less impacts.

\subsection{Comparison with existing eclipse PV impact studies}\label{subsec:Comparism}
The National Renewable Energy Laboratory (NREL) conducted a pre-event analysis of the August 21, 2017 eclipse on the Western Electricity Coordinating Council (WECC) \cite{veda2018evaluating}. The loss in generation was estimated to be around $5.2$ GW of which $4$ GW loss was from utility scale PV and the rest from the distributed PV systems. The post event analysis  based on the data collected by NREL researchers showed that loss in utility scale PV generation was accurately predicted. There were no major impacts on the reliability and and the power quality of the system. Conventional power sources such as  natural gas generators and hydro power were ramped up to mitigate the impacts of the loss in generation form the PVs. The study provided some insights to how more accurate future solar eclipse impact studies can be carried out. With increasing level of penetration, larger system-wide disturbances and impact is expected. Also, a post August 21, 2017 eclipse study was carried out by authors of \cite{8521057}. The paper used the solar irradiance and temperature measurements from the solar eclipse events, to simulate a real-time grid connected utility scale PV. The study provided some insights on the spinning reserve or sizes of storages that can be dispatched during the solar eclipse, the necessity for coordination of these PV and energy storage systems in other to maintain the system's stability as well as the importance of fast frequency control during this event. Load control methods were also discussed as effective alternative methods to address the impact of the solar eclipse event. Most of the case studies presented after the event were not at the device level. PV site performance, detailed power quality analysis, and voltage regulating device operations were not reported in details as presented in this paper.

\section{Conclusion and Future Work} \label{sec:conclusion}
When the penetration of distributed PV systems into the smart grid increases, natural phenomena such as solar eclipse would have significant impacts on the systems and the larger grid. To demonstrate the impacts, this paper explored how the eclipse of August 21, 2017 impacted the performance of two PV systems, $\mathcal{A}$ located in Miami and $\mathcal{B}$ located in Daytona. It was observed that System $\mathcal{A}$, despite being larger in generation capacity than $\mathcal{B}$ showed a slower increase in performance during the eclipse peak owing to a slower drop in irradiance, ambient temperature and module temperature. The steeper drop in these parameters for System $\mathcal{B}$ showed a pronounced effect in the PPI increase for it. Further statistical analysis showed the strong positive relationship between these three parameters both within and across the two systems. This relationship can be expanded to more number of systems over different geographical areas to enable aggregation studies to meet dynamic demands during such predictable events.

This paper investigated the effects of partial solar eclipse on power quality parameters, such as, average RMS voltage, harmonics, and flicker at POI of System $\mathcal{A}$ and presented how those parameters vary during eclipse period. The variability of system power indices were found to be within allowable limits in accordance to IEEE Standard. The results show that System $\mathcal{A}$ had minimum impact on power quality metrics at current penetration level during the eclipse but could have severe effect at high penetration scenarios.

The impact of the eclipse on the feeder was studied by modeling the real parameters of the two feeders ($\mathcal{A}$ and $\mathcal{B}$) as well as the PV systems ($\mathcal{A}$ and $\mathcal{B}$) into a standard IEEE 8500 test distribution network. The simulation results showed the various impacts of the eclipse on the voltage profile as well as operations of the voltage regulation devices in the network. The voltage regulators and the capacitor banks closest to the locations of the PV were forced to operate more frequently during the severe ramp down caused by the eclipse event. The PV system $\mathcal{B}$ , which is close to the substation had little impact on the voltage profile while PV system $\mathcal{B}$ which is further downstream from the substation caused and an increase in the voltage at the POI during the eclipse event. The losses in the network was also impacted by the eclipse event. The ramping in the PV power output led to an increase in the overall system losses during the eclipse.

Finally, the impact of the eclipse at a wider, management-area level, was analyzed by first understanding the relationship between common weather parameters and reliability indices using the regression models. Taking the derived regression models as inputs, we proposed an MLP to forecast the daily reliability indices using time series of common weather data. In addition, we can derive the sensitivity of each common weather parameter with respective to the daily numbers of sustainable interruptions. For the utility management area in Florida, we can find that the lightning strike is the most important common weather parameter impacting on the reliability performance of the smart grid distribution networks, while the average temperature has the least impacts.

As a future work, a PV generation forecasting model will be developed to explore how the utilities can predict the generation profiles of PV systems of different sizes and at penetration levels, which could help in the planning processes. Factors influencing the performance of the estimation model will also be investigated. Further, additional statistical methods such as regression and dependency analysis will be conducted in addition to correlation and relationship study to explore aggregation opportunities during eclipse to manage surges or drops in loads. This study was done without any voltage and frequency regulation from the smart inverters. With increasing PV penetration, the use of smart inverters for voltage and frequency regulation will become inevitable. The use of the various smart inverter functionalities to address the various challenges that will be imposed during an eclipse events will be investigated. The optimal settings and coordination of this devices during these events will be analyzed. The economic consequence of possible reduction in life span of these legacy devices due to the increase in their switching will also be of interest to many utility companies. Finally, for the reliability analysis, other factors such as power system equipment failure rates and aging of distribution network components will be considered in addition to the weather parameters for the regression and prediction methods discussed in this paper.

\section*{Acknowledgments}
The material published is a result of the research supported by the National Science Foundation under the Award number CNS-1553494.




\begin{thebibliography}{10}
\providecommand{\url}[1]{#1}
\csname url@samestyle\endcsname
\providecommand{\newblock}{\relax}
\providecommand{\bibinfo}[2]{#2}
\providecommand{\BIBentrySTDinterwordspacing}{\spaceskip=0pt\relax}
\providecommand{\BIBentryALTinterwordstretchfactor}{4}
\providecommand{\BIBentryALTinterwordspacing}{\spaceskip=\fontdimen2\font plus
\BIBentryALTinterwordstretchfactor\fontdimen3\font minus
  \fontdimen4\font\relax}
\providecommand{\BIBforeignlanguage}[2]{{%
\expandafter\ifx\csname l@#1\endcsname\relax
\typeout{** WARNING: IEEEtran.bst: No hyphenation pattern has been}%
\typeout{** loaded for the language `#1'. Using the pattern for}%
\typeout{** the default language instead.}%
\else
\language=\csname l@#1\endcsname
\fi
#2}}
\providecommand{\BIBdecl}{\relax}
\BIBdecl

\bibitem{Hosenuzzaman2014}
M.~Hosenuzzaman, N.~Rahim, J.~Selvaraj, and M.~Hasanuzzaman, ``{Factors
  affecting the PV based power generation},'' \emph{IET Seminar Digest}, vol.
  2014, no. CP659, 2014.

\bibitem{adityasysjournal}
D.~Saleem, A.~Sundararajan, A.~Sanghvi, J.~Rivera, A.~I. Sarwat, and
  B.~Kroposki, ``{A Multidimensional Holistic Framework for the Security of
  Distributed Energy and Control Systems},'' \emph{IEEE Systems Journal}, 2019.

\bibitem{protocollevel2018}
A.~Sundararajan, A.~Chavan, D.~Saleem, and A.~I. Sarwat, ``A survey of
  protocol-level challenges and solutions for distributed energy resource
  cyber-physical security,'' \emph{MDPI Energies}, no.~9, p. 2360, September
  2018.

\bibitem{theft2017}
L.~Wei, A.~Sundararajan, A.~Sarwat, S.~Biswas, and E.~Ibrahim, ``A distributed
  intelligent framework for electricity theft detection using benford's law and
  stackelberg game,'' in \emph{Resilience Week}, Sep 2017, pp. 5--11.

\bibitem{iotbookchap}
A.~I. Sarwat, A.~Sundararajan, and I.~Parvez, ``Trends and future directions of
  research for smart grid iot sensor networks,'' \emph{{Proceedings of
  International Symposium on Sensor Networks, Systems and Security}}, May 2018.

\bibitem{CAISO2017}
{California Independent System Operator}, ``{Performance of ISO's system during
  August 21, 2017 Eclipse},'' p.~12, 2017.

\bibitem{wide2017}
A.~Uttam and Z.~Hongming, ``{A Wide-Area Perspective on the August 21 , 2017
  Total Solar Eclipse, White Paper},'' no. April, 2017.

\bibitem{en11071782}
\BIBentryALTinterwordspacing
T.~O. Olowu, A.~Sundararajan, M.~Moghaddami, and A.~I. Sarwat, ``{Future
  Challenges and Mitigation Methods for High Photovoltaic Penetration: A
  Survey},'' \emph{Energies}, vol.~11, no.~7, 2018. [Online]. Available:
  \url{http://www.mdpi.com/1996-1073/11/7/1782}
\BIBentrySTDinterwordspacing

\bibitem{nreleclipse2018}
S.~Veda, Y.~Zhang, J.~Tan, J.~Duckworth, N.~Gilroy, D.~Hettinger, S.~Ericson,
  J.~Ausmus, S.~Kincic, X.~Zhang, and G.~Yuan, ``Evaluating the impact of the
  2017 solar eclipse on the u.s. western interconnection operations,''
  \emph{{U.S. Department of Energy Solar Energy Technologies Office Technical
  Report}}, 2018.

\bibitem{nrelreportaditya}
\BIBentryALTinterwordspacing
Z.~Peterson, M.~Coddington, F.~Ding, B.~Sigrin, D.~Saleem, K.~Horowitz, S.~E.
  Baldwin, B.~Lydic, S.~C. Stanfield, N.~Enbar, S.~Coley, A.~Sundararajan, and
  C.~Schroeder, ``An overview of distributed energy resource (der)
  interconnection: Current practices and emerging solutions,'' \emph{{NREL
  Technical Report (number NREL/TP-6A20-72102)}}, April 2019. [Online].
  Available: \url{https://www.nrel.gov/docs/fy19osti/72102.pdf}
\BIBentrySTDinterwordspacing

\bibitem{duong2018}
M.~Q. {Duong}, N.~T.~N. {Tran}, G.~N. {Sava}, S.~{Leva}, and M.~{Mussetta},
  ``The impact of 150mwp phoan solar photovoltaic project into vietnamese
  quangngai - grid,'' in \emph{2018 International Conference and Exposition on
  Electrical And Power Engineering (EPE)}, Oct 2018, pp. 0498--0502.

\bibitem{adityampce}
A.~Sundararajan, T.~Khan, A.~Moghadasi, and A.~I. Sarwat, ``Survey on
  synchrophasor data quality and cybersecurity challenges, and evaluation of
  their interdependencies,'' \emph{Journal of Modern Power Systems and Clean
  Energy}, pp. 1--19, 2018.

\bibitem{adityaacm}
{A. Sundararajan and A. I. Sarwat and A. Pons}, ``{A Survey on Modality
  Characteristics, Performance Evaluation Metrics, and Security for Traditional
  and Wearable Biometric Systems},'' \emph{{ACM Computing Surveys}}, vol.~52,
  no.~2, pp. 1--35, 2019.

\bibitem{pvperf01}
\BIBentryALTinterwordspacing
{Kurinec, Santosh K.}, {Kucer, Michal}, and {Schlein, Bill}, ``Monitoring a
  photovoltaic system during the partial solar eclipse of august 2017,''
  \emph{EPJ Photovolt.}, vol.~9, p.~7, 2018. [Online]. Available:
  \url{https://doi.org/10.1051/epjpv/2018005}
\BIBentrySTDinterwordspacing

\bibitem{pvperfnrel1}
\BIBentryALTinterwordspacing
C.~Deline, N.~DiOrio, D.~Jordan, and F.~Toor, ``Progress \& frontiers in pv
  performance,'' \emph{Solar Power International}, 2016. [Online]. Available:
  \url{https://www.nrel.gov/docs/fy16osti/67174.pdf}
\BIBentrySTDinterwordspacing

\bibitem{pvperfsunspec}
\BIBentryALTinterwordspacing
J.~Mokri and J.~Cunningham, ``Pv system performance assessment,'' \emph{SunSpec
  Alliance and San Jose State University Technical Report}, 2014. [Online].
  Available:
  \url{https://sunspec.org/wp-content/uploads/2015/06/SunSpec-PV-System-Performance-Assessment-v2.pdf}
\BIBentrySTDinterwordspacing

\bibitem{pvperfnrel2}
\BIBentryALTinterwordspacing
P.~Denholm, J.~Eichman, and R.~Margolis, ``Evaluating the technical and
  economic performance of pv plus storage power plants,'' \emph{A National
  Renewable Energy Laboratory (NREL) Technical Report}, 2017. [Online].
  Available: \url{https://www.nrel.gov/docs/fy16osti/67174.pdf}
\BIBentrySTDinterwordspacing

\bibitem{pvperfnrel3}
\BIBentryALTinterwordspacing
S.~Kurtz, E.~Riley, J.~Newmiller, T.~Dierauf, A.~Kimber, J.~McKee,
  R.~Flottemesch, and P.~Krishnani, ``Analysis of photovoltaic system energy
  performance evaluation method,'' \emph{A National Renewable Energy Laboratory
  (NREL) Technical Report}, 2013. [Online]. Available:
  \url{https://www.nrel.gov/docs/fy14osti/60628.pdf}
\BIBentrySTDinterwordspacing

\bibitem{pvperf4}
\BIBentryALTinterwordspacing
E.~Rhee, ``Not just another day of sun: Reviewing the solar eclipse's effect on
  pv system performance,'' \emph{A Sol Systems Online Article}, 2017. [Online].
  Available: \url{solsystems.com/blog/2017/09/22/not-just-another-day-of
  -sun-reviewing-the-solar-eclipses-effect-on-pv-system/}
\BIBentrySTDinterwordspacing

\bibitem{pvperf5}
M.~Libra, P.~Kourim, and V.~Poulek, ``Behavior of photovoltaic system during
  solar eclipse in prague,'' \emph{International Journal of Photoenergy}, 2016.

\bibitem{pvperf6}
S.~V. Dhople and A.~D. Dominguez-Garcia, ``Estimation of photovoltaic system
  reliability and performance metrics,'' \emph{IEEE Transactions on Power
  Systems}, vol.~27, no.~1, 2011.

\bibitem{pvperf7}
B.~S. Kumar and K.~Sudhakar, ``{Performance evaluation of 10 MW grid connected
  solar photovoltaic power plant in India}, doi= "10.1016/j.egyr.2015.10.001",
  journal = {Elsevier Energy Reports}, year = 2015.''

\bibitem{HAQUE20161195}
M.~M. Haque and P.~Wolfs, ``{A review of high PV penetrations in LV
  distribution networks: Present status, impacts and mitigation measures},''
  \emph{Renewable and Sustainable Energy Reviews}, vol.~62, pp. 1195 -- 1208,
  2016.

\bibitem{7856217}
S.~Ghosh and S.~Rahman, ``Global deployment of solar photovoltaics: Its
  opportunities and challenges,'' in \emph{2016 IEEE PES Innovative Smart Grid
  Technologies Conference Europe (ISGT-Europe)}, Oct 2016, pp. 1--6.

\bibitem{VoltageIssue}
\BIBentryALTinterwordspacing
R.~Seguin, J.~Woyak, D.~Costyk, J.~Hambrick, and B.~Mather, ``{High-Penetration
  PV Integration Handbook for Distribution Engineers},'' \emph{A National
  Renewable Energy Laboratory (NREL) Technical Report}, 2016. [Online].
  Available: \url{https://www.nrel.gov/docs/fy16osti/63114.pdf}
\BIBentrySTDinterwordspacing

\bibitem{6939147}
A.~Chidurala, T.~K. Saha, N.~Mithulananthan, and R.~C. Bansal, ``{}harmonic
  emissions in grid connected pv systems: A case study on a large scale rooftop
  pv site,'' in \emph{2014 IEEE PES General Meeting | Conference Exposition}.

\bibitem{Thesis2013}
A.~Thesis, ``{Impact of Photovoltaic System Penetration on the Operation of
  Voltage Regulator Equipment},'' no. June, 2013.

\bibitem{Veda2018}
\BIBentryALTinterwordspacing
S.~Veda, Y.~Zhang, J.~Tan, E.~Chartan, J.~Duckworth, N.~Gilroy, D.~Hettinger,
  and S.~Ericson, ``{Evaluating the Impact of the 2017 Solar Eclipse},'' 2018.
  [Online]. Available:
  \url{https://www.nrel.gov/news/press/2018/nrel-researchers-measure-impact-of-eclipse-on-electrical-grid.html}
\BIBentrySTDinterwordspacing

\bibitem{WideArea}
``{A Wide-Area Perspective on the August 21 , 2017 Total Solar Eclipse White
  Paper},'' no. April, 2017.

\bibitem{J1}
M.~Panteli and P.~Mancarella, ``Influence of extreme weather and climate change
  on the resilience of power systems: Impacts and possible mitigation
  strategies,'' \emph{Electr. Pow. Syst. Res.}, vol. 127, 2015.

\bibitem{J2}
Y.~Yang, W.~Tang, Y.~Liu, Y.~Xin, and Q.~Wu, ``Quantitative resilience
  assessment for power transmission systems under typhoon weather,'' \emph{IEEE
  Access}, vol.~6, pp. 40\,747--40\,756, 2018.

\bibitem{J3}
M.~Panteli, D.~N. Trakas, P.~Mancarella, and N.~D. Hatziargyriou, ``Boosting
  the power grid resilience to extreme weather events using defensive
  islanding,'' \emph{IEEE Trans. Smart Grid}, vol.~7, no.~6, Nov. 2016.

\bibitem{Arif1}
A.~I. Sarwat, A.~Domijan, M.~H. Amini, A.~Damnjanovic, and A.~Moghadasi,
  ``Smart grid reliability assessment utilizing boolean driven markov process
  and variable weather conditions,'' in \emph{2015 North American Power
  Symposium (NAPS)}, Oct. 2015, pp. 1--6.

\bibitem{Arif2}
A.~I. Sarwat., M.~Amini, A.~Domijan, A.~Damnjanovic, and F.~Kaleem,
  ``Weather-based interruption prediction in the smart grid utilizing
  chronological data,'' \emph{J. Mod. Pow. Syst. Cl. Ener.}, vol.~4, no.~2, pp.
  308--315, Apr. 2016.

\bibitem{pvsys1}
A.~Anzalchi, A.~Sundararajan, A.~Moghadasi, and A.~I. Sarwat, ``Power quality
  and voltage profile analyses of high penetration grid-tied photovoltaics: A
  case study,'' \emph{IEEE Industry Applications Society Annual Meeting}, 2017.

\bibitem{missingdata}
A.~{Sundararajan} and A.~I. {Sarwat}, ``{Evaluation of Missing Data Imputation
  Methods for an Enhanced Distributed PV Generation Prediction},'' in
  \emph{{2019-2020 Advances in Intelligent Systems and Computing}}, In Press.

\bibitem{pvsys2}
A.~Sundararajan and A.~I. Sarwat, ``Roadmap to prepare distribution grid-tied
  photovoltaic site data for performance monitoring,'' \emph{International
  Conference on Big Data, IoT and Data Science}, 2017.

\bibitem{fogbookchap}
A.~Anzalchi, A.~Sundararajan, L.~Wei, A.~Moghadasi, and A.~I. Sarwat, ``Future
  directions to the application of distributed fog computing in smart grid
  systems,'' \emph{{Cloud Security: Concepts, Methodologies, Tools, and
  Applications}}, January 2019.

\bibitem{Arritt2010}
R.~F. Arritt and R.~C. Dugan, ``{The IEEE 8500-node test feeder},'' \emph{2010
  IEEE PES Transmission and Distribution Conference and Exposition: Smart
  Solutions for a Changing World}, pp. 1--6, 2010.

\bibitem{performancemetrics}
B.~Marion, J.~Adelstein, K.~Boyle, H.~Hayden, B.~Hammond, T.~Fletcher,
  B.~Canada, D.~Narang, D.~Shugar, H.~Wenger, A.~Kimber, L.~Mitchell, G.~Rich,
  and T.~Townsend, ``{Performance Parameters for Grid-Connected PV Systems},''
  \emph{31st IEEE Photovoltaics Specialists Conference and Exhibition}, 2005.

\bibitem{performancemetrics1}
\BIBentryALTinterwordspacing
S.~Pless, M.~Deru, P.~Torcellini, and S.~Hayter, ``Procedure for measuring and
  reporting the performance of photovoltaic systems in buildings,'' \emph{A
  National Renewable Energy Laboratory (NREL) Technical Report}, 2005.
  [Online]. Available: \url{https://www.nrel.gov/docs/fy06osti/38603.pdf}
\BIBentrySTDinterwordspacing

\bibitem{performancemetrics2}
A.~Haibaoui, B.~Hartiti, A.~Elamim, M.~Karami, and A.~Ridah, ``{Performance
  Indicators For Grid-Connected PV Systems: A Case Study In Casablanca,
  Morocco},'' \emph{{IOSR Journal of Electrical and Electronics Engineering
  (IOSR-JEEE)}}, pp. 55--65, 2017.

\bibitem{performancemetrics3}
T.~Townsend, C.~Whitaker, B.~Farmer, and H.~Wenger, ``{A new performance index
  for PV system analysis},'' \emph{Proceedings of IEEE 1st World Conference on
  Photovoltaic Energy Conversion - WCPEC (A Joint Conference of PVSC, PVSEC and
  PSEC)}, 1994.

\bibitem{derate}
\BIBentryALTinterwordspacing
E.~Energy, ``{Guide to PVWatts Derate Factors for Enphase Systems When Using PV
  System Design Tools},'' \emph{An Enphase Energy Technical Report}, 2014.
  [Online]. Available:
  \url{https://enphase.com/sites/default/files/Enphase_PVWatts_Derate_Guide_ModSolar_06-2014.pdf}
\BIBentrySTDinterwordspacing

\bibitem{derate1}
B.~Yerli, M.~K. Kaymak, E.~İzgi, A.~Öztopal, and A.~D. Şahin, ``{Effect of
  Derating Factors on Photovoltaics under Climatic Conditions of Istanbul},''
  \emph{World Academy of Science, Engineering and Technology}, 2010.

\bibitem{correctedpr}
\BIBentryALTinterwordspacing
T.~Dierauf, A.~Growitz, S.~Kurtz, J.~L.~B. Cruz, B.~Riley, and C.~Hansen,
  ``Weather-corrected performance ratio,'' \emph{A National Renewable Energy
  Laboratory (NREL) Technical Report}, 2013. [Online]. Available:
  \url{https://www.nrel.gov/docs/fy13osti/57991.pdf}
\BIBentrySTDinterwordspacing

\bibitem{correctedpr1}
R.~Bohra, ``{Performance Analysis of 1MW SPV Plant; Temperature Corrected
  PR},'' \emph{Solar Power Energetica India Article}, 2014.

\bibitem{correctedpr2}
H.~A. Basson and J.~C. Pretorius, ``Risk mitigation of performance ratio
  guarantees in commercial photovoltaic systems,'' \emph{International
  Conference on Renewable Energies and Power Quality}, 2016.

\bibitem{CHAUDHARY20183279}
\BIBentryALTinterwordspacing
P.~Chaudhary and M.~Rizwan, ``{Voltage regulation mitigation techniques in
  distribution system with high PV penetration: A review},'' \emph{Renewable
  and Sustainable Energy Reviews}, vol.~82, pp. 3279 -- 3287, 2018. [Online].
  Available:
  \url{http://www.sciencedirect.com/science/article/pii/S1364032117313989}
\BIBentrySTDinterwordspacing

\bibitem{1242609}
T.~Hoevenaars, K.~LeDoux, and M.~Colosino, ``{Interpreting IEEE STD 519 and
  meeting its harmonic limits in VFD applications},'' in \emph{IEEE Industry
  Applications Society 50th Annual Petroleum and Chemical Industry Conference,
  2003. Record of Conference Papers.}, Sept 2003, pp. 145--150.

\bibitem{FlickerMeasurement}
\BIBentryALTinterwordspacing
A.~L. J.J.~Gutierrez, J.~Ruiz and L.~Leturiondo, ``Measurement of voltage
  flicker: Application to grid-connected wind turbines,'' 2010. [Online].
  Available: \url{http://cdn.intechopen.com/pdfs/9948/InTech}
\BIBentrySTDinterwordspacing

\bibitem{8600542}
M.~{Jafari}, T.~O. {Olowu}, and A.~I. {Sarwat}, ``{Optimal Smart Inverters
  Volt-VAR Curve Selection with a Multi-Objective Volt-VAR Optimization using
  Evolutionary Algorithm Approach},'' in \emph{2018 North American Power
  Symposium (NAPS)}, Sep. 2018, pp. 1--6.

\bibitem{8600557}
T.~O. {Olowu}, M.~{Jafari}, and A.~I. {Sarwat}, ``{A Multi-Objective
  Optimization Technique for Volt-Var Control with High PV Penetration using
  Genetic Algorithm},'' in \emph{{2018 North American Power Symposium (NAPS)}},
  Sep. 2018, pp. 1--6.

\bibitem{Stiles2005}
J.~Stiles, ``{Voltage Regulators},'' no. June, pp. 8--10, 2005.

\bibitem{Manbachi2015}
M.~Manbachi, ``{Smart Grid Adaptive Volt-VAR Optimization in Distribution
  Networks},'' 2015.

\bibitem{Harrell2015}
F.~E. Harrell, \emph{Ordinal Logistic Regression}.\hskip 1em plus 0.5em minus
  0.4em\relax Cham: Springer International Publishing, 2015, pp. 311--325.

\bibitem{wei2018hybrid}
L.~Wei and A.~I. Sarwat, ``Hybrid integration of multilayer perceptrons and
  parametric models for reliability forecasting in the smart grid,''
  \emph{arXiv preprint arXiv:1810.05004}, 2018.

\bibitem{itec}
A.~{Khalid}, A.~{Sundararajan}, I.~{Acharya}, and A.~I. {Sarwat}, ``{Prediction
  of Li-Ion Battery State of Charge Using Multilayer Perceptron and Long
  Short-Term Memory Models},'' in \emph{2019 IEEE Transportation
  Electrification Conference (ITEC)}, In Press.

\bibitem{eeeic}
A.~{Khalid}, A.~{Sundararajan}, and A.~I. {Sarwat}, ``{A Multi-Step Predictive
  Model to Estimate Li-Ion State of Charge for Higher C-Rates},'' in
  \emph{{2019 IEEE International Conference on Environment and Electrical
  Engineering and 2019 IEEE Industrial and Commercial Power Systems Europe
  (EEEIC / I\&CPS Europe)}}, In Press.

\bibitem{xavierdist}
X.~Glorot and Y.~Bengio, ``Understanding the difficulty of training deep
  feedforward neural networks,'' in \emph{13th International Conference on
  Artificial Intelligence and Statistics (AISTATS)}, 2010, pp. 1--8.

\bibitem{uniformdist}
D.~Nguyen and B.~Widrow, ``Improving the learning speed of 2-layer neural
  networks by choosing initial values of the adaptive weights,'' in \emph{1990
  IJCNN International Joint Conference on Neural Networks}, June 1990, pp.
  21--26 vol.3.

\bibitem{sigmoidal}
B.~Karlik and A.~V. Olgac, ``{Performance Analysis of Various Activation
  Functions in Generalized MLP Architectures of Neural Networks},''
  \emph{International Journal of Artificial Intelligence And Expert Systems
  (IJAE)}, no.~4, pp. 111--122, 2011.

\bibitem{hirose1991back}
Y.~Hirose, K.~Yamashita, and S.~Hijiya, ``Back-propagation algorithm which
  varies the number of hidden units,'' \emph{Neural Networks}, vol.~4, no.~1,
  pp. 61--66, 1991.

\bibitem{patsalides2007effect}
M.~Patsalides, D.~Evagorou, G.~Makrides, Z.~Achillides, G.~E. Georghiou,
  A.~Stavrou, V.~Efthimiou, B.~Zinsser, W.~Schmitt, and J.~H. Werner, ``The
  effect of solar irradiance on the power quality behaviour of grid connected
  photovoltaic systems,'' in \emph{International Conference on Renewable Energy
  and Power Quality}, 2007, pp. 1--7.

\bibitem{Hybrid1}
Y.-C. Lai, Y.-A. Huang, and H.-Y. Chu, ``Estimation of rail capacity using
  regression and neural network,'' \emph{Neural Comput. Appl.}, vol.~25, no.~7,
  pp. 2067--2077, Dec 2014.

\bibitem{veda2018evaluating}
S.~Veda, Y.~Zhang, J.~Tan, E.~K. Chartan, N.~Gilroy, D.~J. Hettinger, S.~J.
  Ericson, J.~Ausmus, S.~Kincic, X.~Zhang \emph{et~al.}, ``{Evaluating the
  Impact of the 2017 Solar Eclipse on U.S. Western Interconnection
  Operations},'' National Renewable Energy Lab.(NREL), Golden, CO (United
  States), Tech. Rep., 2018.

\bibitem{8521057}
A.~{Arzani}, I.~{Jayawardene}, P.~{Arunagirinathan}, and G.~K.
  {Venavagamoorthy}, ``Impact of solar eclipse on utility grid operations,'' in
  \emph{2018 IEEE PES/IAS PowerAfrica}, June 2018, pp. 1--9.

\end{thebibliography}

\end{document}